# Hayabusa2 Extended Mission: New Voyage to Rendezvous with a Small Asteroid Rotating with a Short Period

M. Hirabayashi[1], Y. Mimasu[2], N. Sakatani[3], S. Watanabe[4], Y. Tsuda[2], T. Saiki[2], S. Kikuchi[2], T. Kouyama[5], M. Yoshikawa[2], S. Tanaka[2], S. Nakazawa[2], Y. Takei[2], F. Terui[2], H. Takeuchi[2], A. Fujii[2], T. Iwata[2], K. Tsumura[6], S. Matsuura[7], Y. Shimaki[2], S. Urakawa[8], Y. Ishibashi[9], S. Hasegawa[2], M. Ishiguro[10], D. Kuroda[11], S. Okumura[8], S. Sugita[12], T. Okada[2], S. Kameda[3], S. Kamata[13], A. Higuchi[14], H. Senshu[15], H. Noda[16], K. Matsumoto[16], R. Suetsugu[17], T. Hirai[15], K. Kitazato[18], D. Farnocchia[19], S.P. Naidu[19], D.J. Tholen[20], C.W. Hergenrother[21], R.J. Whiteley[22], N. A. Moskovitz[23], P.A. Abell[24], and the Hayabusa2 extended mission study group.

[1]Auburn University, Auburn, AL, USA (thirabayashi@auburn.edu)
[2]Japan Aerospace Exploration Agency, Kanagawa, Japan
[3]Rikkyo University, Tokyo, Japan
[4]Nagoya University, Aichi, Japan
[5]National Institute of Advanced Industrial Science and Technology, Tokyo, Japan
[6]Tokyo City University, Tokyo, Japan
[7]Kwansei Gakuin University, Hyogo, Japan
[8]Japan Spaceguard Association, Okayama, Japan
[9]Hosei University, Tokyo, Japan
[10]Seoul National University, Seoul, South Korea
[11]Kyoto University, Kyoto, Japan
[12]University of Tokyo, Tokyo, Japan
[13]Hokkaido University, Hokkaido, Japan
[14]University of Occupational and Environmental Health, Fukuoka, Japan
[15]Chiba Institute of Technology, Chiba, Japan
[16]National Astronomical Observatory of Japan, Iwate, Japan
[17]National Institute of Technology, Oshima College, Yamaguchi, Japan
[18]University of Aizu, Fukushima, Japan
[19]Jet Propulsion Laboratory, California Institute of Technology, Pasadena, CA, USA
[20]University of Hawai'i, Manoa, HI, USA
[21]University of Arizona, Tucson, AZ, USA
[22]Asgard Research, Denver, CO, USA
[23]Lowell Observatory, Flagstaff, AZ, USA
[24]NASA Johnson Space Center, Houston, TX, USA

**Highlights**

1. Hayabusa2 plans its extended mission to rendezvous with 1998 KY26 in 2031.

2. We show science assessments to select the final two candidates, 1998 KY26 and 2001 AV43.

3. The mission will explore the origin and evolution of small bodies and planetary defense.

**Abstract**

Hayabusa2 is the Japanese Asteroid Return Mission and targeted the carbonaceous asteroid Ryugu, conducted by the Japan Aerospace Exploration Agency (JAXA). The goal of this mission was to conduct proximity operations including remote sensing observations, material sampling, and a Small Carry-On Impact experiment, as well as sample analyses. As of September 2020, the spacecraft is on the way back to Earth with samples from Ryugu with no critical issues after the successful departure in November 2019. Here, we propose an extended mission in which the spacecraft will rendezvous with a small asteroid with ~30 m - ~40 m in diameter that is rotating at a spin period of ~10 min after an additional ~10-year cruise phase. We introduce that two scenarios are suitable for the extended mission. In the first scenario, the spacecraft will perform swing-by maneuvers at Venus once and Earth twice to arrive at asteroid 2001 AV43. In the second scenario, it will perform swing-by maneuvers at Earth twice to reach asteroid 1998 KY26. In both scenarios, the mission will continue until the early 2030s. JAXA recently released the decision that the spacecraft will rendezvous with 1998 KY26. This paper focuses on our scientific assessments of the two scenarios but leaves the decision process to go to 1998 KY26 for future reports. Rendezvous operations will be planned to detail the physical properties and surrounding environments of the target, one of the smallest elements of small planetary bodies. By achieving the planned operations, the mission will provide critical hints on the violent histories of collisions and accumulations of small bodies in the solar system. Furthermore, the established scientific knowledge and techniques will advance key technologies for planetary defense.

# 1. Introduction

## 1.1. Brief summary of nominal mission

Hayabusa2 is the second asteroid sample return mission led by the Japan Aerospace Exploration Agency (JAXA). Since launched in 2014, the spacecraft has successfully arrived at (162173) Ryugu in June 2018, conducted detailed scientific investigations, and departed from this asteroid in November 2019 to return samples to Earth. Spacecraft operations included remote sensing mapping, material sampling, surface observations using robotic rovers and landers, and an impact experiment, which revealed Ryugu's environment. The spacecraft is equipped with the following remote sensing instruments (Watanabe et al., 2017): Optical Navigation Camera (ONC) (Kameda et al., 2017, 2015; Suzuki et al., 2018; Tatsumi et al., 2019); Laser Altimeter (LIDAR) (Mizuno et al., 2017; Senshu et al., 2017; Yamada et al., 2017); Near-Infrared Spectrometer (NIRS3) (Iwata et al., 2017); and Thermal Infrared Imager (TIR) (Arai et al., 2017; Okada et al., 2017; Takita et al., 2017). The spacecraft also had multiple rovers and a lander that includes multiple remote sensing instruments (Bibring et al., 2017; Grott et al., 2017; Herčík et al., 2017; Ho et al., 2017; Jaumann et al., 2017), a sampling system (Okazaki et al., 2017; Sawada et al., 2017b), a Small Carry-on Impactor (SCI) (Saiki et al., 2020), and a Deployable Camera (DCAM3) (Ishibashi et al., 2017; Ogawa et al., 2017; Sawada et al., 2017a) for observations of the SCI impact crater formation (Arakawa et al., 2017). The major critical operations, including the material sampling and SCI experiment, were successful during the proximity operation phase.

While staying at Ryugu after the arrival in June 2018, the spacecraft conducted pioneering scientific investigations that brought key understandings of Ryugu's environment. Below, we summarize the main scientific findings made by Hayabusa2 as of September 2020 while many exciting outcomes including material analyses are coming out at present.

1. Weak (but non-zero) absorption at a spectral band of 2.7 $\mu$m shows the existence of hydrated minerals (Kitazato et al., 2019), and the observed flat spectra and extremely low reflectance from visible to near-infrared is consistent with thermally metamorphosed carbonaceous chondrites and main-belt asteroid families, Polana and Eulalia (Barucci et al., 2019; Galiano et al., 2020; Pilorget et al., 2020; Sugita et al., 2019; Tatsumi et al., 2020a).

2. Observations of surface features, such as impact craters (Hirata et al., 2020), impactor fragments (Tatsumi et al., 2020b), mass wasting (Sugita et al., 2019), and spectral slope distribution (Barucci et al., 2019; Galiano et al., 2020), led to a scenario of the evolution of Ryugu through parent-body processes including collisional disruption and accumulation and orbital/spin transition (Morota et al., 2020; Sugita et al., 2019).

3. Ryugu's top shape (a spheroidal shape with an equatorial ridge) implies the violent formation process driven by reaccumulation of fragments after catastrophic disruption of parent bodies and the following evolution process due to fast rotation (Hirabayashi et al., 2020, 2019b; Michel et al., 2020a; Watanabe et al., 2019).

4. Thermal signatures of Ryugu's surface suggest the highly porous nature of boulders (Grott et al., 2019; Okada et al., 2020; Shimaki et al., 2020), significantly contributing to the low bulk density, 1.19 g/cm$^3$ (Watanabe et al., 2019).

5. Close observations on the surface revealed the detailed characteristics of boulders and regolith, which are consistent with those seen in carbonaceous chondrite meteorites, as well as the magnetic environment on the surface (Grott et al., 2019; Hercik et al., 2020; Jaumann et al., 2019; Scholten et al., 2019).

6. Demonstration of a Small Carry-on Impactor experiment showed that the impact formation process on Ryugu could be characterized in the gravity-dominant regime, implying that Ryugu's top surface would consist of a mechanically weak structure (Arakawa et al., 2020; Saiki et al., 2020). This also constrained the crater retention age on Ryugu, revealing that Ryugu's major topography such as equatorial ridges was formed ~10$^7$ years ago (Morota et al., 2020).

### *1.2. Transition to extended mission*

After the planned proximity operations were completed with no critical issues, the spacecraft successfully departed from Ryugu in November 2019. The spacecraft is planned to arrive at Earth with materials sampled at two locations on this asteroid and release the reentry capsule in December 2020. The nominal mission completes its operations after the release. Based on our assessment that the onboard systems are nominal, and the fuel remains enough for another mission, we seek another opportunity to visit other small bodies. The spacecraft that is already flying offers a crucial chance to give further constraints to the latest findings from the nominal mission (Section 1.1) to better understand the correlations between small bodies and the solar system evolution and further explore unsolved scientific questions related to planetary defense. We define the extended mission as any spacecraft operations after the arrival at Earth with the release of the reentry capsule. This manuscript reports our preliminary scientific assessments of the Hayabusa2 extended mission. We plan to report detailed technical assessments in the future.

We have elaborated two mission scenarios. The first mission scenario is to perform swing-by maneuvers at Venus once and Earth twice after the departure from Earth in 2020. After 10 years of cruising, the spacecraft rendezvouses with asteroid 2001 AV43, which has a diameter of ~40 m and a spin period of ~10 min (Whiteley et al., 2002). We define this scenario as the EVEEA scenario, where E, V, and A stand for Earth, Venus, and Asteroid, respectively. The second scenario is to perform swing-by maneuvers at Earth twice to rendezvous with asteroid 1998 KY26, which has a diameter of ~30 m and a spin period of ~10 min (Ostro et al., 1999). The cruise phase continues for ~10 years for the spacecraft to arrive at this asteroid. In this scenario, the spacecraft may make a fly-by encounter of asteroid 2001 CC21, which has a diameter of ~700 m, in 2026. We call this scenario the E(A)EEA scenario. With further assessments of engineering constraints, as well as scientific significances, the Hayabusa2 extended mission team made a final decision for which scenario would be selected as the official extended mission schedule. JAXA released our decision that the E(A)EEA scenario is the official mission scenario in September 2020. In this paper, we focus on the selection process to define the two scenarios. The details about the decision to select the E(A)EEA scenario will be reported in the future. We finally note that an extension plan was also reported recently by Sarli and Tsuda (2017); however, our mission analyses include detailed engineering and scientific criteria, which gave different options.

Here is the outline of this work. Section 2 briefly summaries the remote sensing instruments onboard the spacecraft that will play key roles in the extended mission. Section 3 discusses the scientific motivation of the Hayabusa2 extended mission. Section 4 defines the mission objectives. Section 5 shows the selection process of the target candidates. Section 6 introduces the proposed mission scenarios, given the target candidates. Section 7 shows the target candidates for rendezvous operations. Finally, Section 8 gives potential observations during the long-term cruise phase.

## 2. Scientific instruments onboard Hayabusa2

This section briefly reviews the remote sensing instruments onboard the Hayabusa2 extended mission, which will be useful during the extended mission. There are four instruments: Optical Navigation Camera (ONC); Thermal Infrared Imager (TIR); Near Infrared Spectrometer (NIRS3); Laser Altimeter (LIDAR). In addition to these instruments, there are one projectile that was supposed to be used for sampling and one target marker. The potential use of the projectile and target marker is currently under discussion and beyond the scope of this paper.

### 2.1. Optical Navigation Camera (ONC)

The Optical Navigation Camera suite consists of one telescopic camera (ONC-T) and two wide-angle view cameras (ONC-W1/W2) (Suzuki et al., 2018). The size of images from these imagers is 1,024 pix x 1,024 pix. The field of view (FOV) of ONC-W1/W2 is 69 deg x 69 deg (Instantaneous FOV, or IFOV, of 0.067 deg). The FOV of ONC-T is 6.27 deg x 6.27 deg (IFOV of 0.0061 deg); for example, given a distance of 100 m, we can observe a 11 m x 11 m area. The exposure time of these cameras can be set in multiple ways. The exposure time of ONC-W1 ranges from 0 s to 5.57 s, while that of ONC-W2 is between 0 s and 44.6 s. ONC-T's exposure time can be controlled in the range between 0 s to 178 sec (Kameda et al., 2017). ONC-T has a wheel system that rotates seven band-pass filters and one panchromatic glass window. The center wavelengths of the filters are 0.39 $\mu$m (ul-band), 0.48 $\mu$m (b-band), 0.55 $\mu$m (v-band), 0.59 $\mu$m (Na), 0.70 $\mu$m (x-band), 0.86 $\mu$m (w-band), and 0.95 $\mu$m (p-band). The selection of these wavelengths was due to observations of Ryugu's carbonaceous compositions (Kameda et al., 2017). Particularly, 0.70 $\mu$m was chosen to observe hydrated minerals including serpentine (Kameda et al., 2017, 2015). No critical issues have been reported. However, there are three minor issues.

The first issue is the increase of the occurrence of hot pixels on CCD (Kouyama et al., 2020; Tatsumi et al., 2019) It was confirmed that the occurrence of hot pixels at the time of the departure from Ryugu (November 2019) was 4% higher than the original condition. This observation predicts that the number of hot pixels would increase by 8% or more when the spacecraft arrives at the target in the early 2030s. This issue can be mitigated by keeping the CCD temperature low and the shutter speed short. However, some imaging processes that require longer shutter speeds, such as calibration using stars and exoplanet observations (see Section 8.4), may experience considerable influences due to this issue. The potential solution may be to move the field of view slightly to take multiple images of target objects and eliminate the hot pixel effects by data processing.

The second issue is the lifetime of the filter wheel of ONC-T, which depends on the number of wheel rotation (Kouyama et al., 2020). Prior to the second touchdown on Ryugu, the wheel rotation was set to remain below 1000 times based on ground tests. However, the wheel has kept normal since the completion of the proximity operations. Based on its good performance, therefore, we redefined ~2000 times as the next upper rotation limit. At present, the total rotation reaches ~1500 times. The rest of available rotation (~500 times) needs to be preserved for both necessary observations and calibrations. To minimize the wheel rotation, we plan to fix it and use the v-band filter. If any issue arises with ONC-T, we plan to use ONC-W1/W2 as the alternatives.

The third issue was dust contamination. The spacecraft conducted two touchdowns on Ryugu. During these operations, small dust particles were ejected from the surface of Ryugu (Morota et al., 2020). We confirmed that small dust particles were attached to the ND filter and the CCD cover on ONC-W1 (Kouyama et al., 2020; Morota et al., 2020). While ONC-T also had significant degradation of the sensitivity owing to dust contamination (~15% over the entire rendezvous phase), post-touchdown observations have shown that its relative spectral sensitivity can be calibrated better than 1% (Kouyama et al., 2020). We plan to monitor how this degradation changes during the extended mission.

## 2.2. Thermal Infrared Imager (TIR)

TIR's detector is an uncooled bolometer array, NEC 320A (anti-reflection coating). The size of images from TIR is 344 pix x 260 pix (effective 328 pix x 248 pix). The FOV is 16.7 deg x 12.7 deg, and the IFOV is 0.051 deg. The frame rate is $1/60 - 2.1$ second per frame, depending on image summation. It compresses 32-bit images to 15-bit images. The covered wavelength ranges between 8 $\mu$m and 12 $\mu$m. Given the preflight tests (Okada et al., 2017), TIR's well calibrated temperature range is $233 - 423$ K, and its detectable range may be $150 - 460$ K.

The effective lifetime of TIR is determined by how many times mechanical shutters are operated. TIR's Engineering Model demonstrated 600,000 times. At present, the total times of shuttering is 69,000 times, there are still 500,000 times available. It was confirmed that when the spacecraft departed from Ryugu, TIR was reported to have an increase of the background level up to $15 - 20$ DN, which is equivalent to ~2 K for black-body radiation at 300 K, compared to the level when the spacecraft was launched in 2014. A similar increase also appears in the data of the Longwave Infrared Camera (LIR) onboard Akatsuki, which has the same configuration as TIR (Kouyama et al., 2019). Also, it is possible that dust contamination of the lens during the sampling operations has contributed to the increase. However, this degradation level was expected, and we concluded that it would not affect the extended mission's operations and sciences after the calibration of the background level.

## 2.3. Near Infrared Spectrometer (NIRS3)

The Near Infrared Spectrometer NIRS3 performs near-infrared spectroscopic observations over the effective wavelength range from 1.8 to 3.2 $\mu m$ (Iwata et al., 2017). NIRS3 was designed to have a spectral sampling resolution of 18 nm and achieve a signal-to-noise ratio of >300 at 20 km altitude (from Ryugu) and >50 at 1 km altitude. The spatial resolution is 40 m at 20 km altitude and 2 m at 1 km altitude. The wavelength range covers the primary absorption bands of silicates (1.9 to 2.3 $\mu m$), phyllosilicates (1.9, 2.2-2.3, 2.7, and 2.9-3.2 $\mu m$) and carbonates (2.3, 2.5, and 2.8 $\mu m$). NIRS3 consists of the Spectrometer Unit (NIRS3-S), the Analogue Electronics Unit (NIRS3-AE), and the Harness Cable (NIRS3-HNS). The linear image sensor consists of 128-element indium-arsenide (InAs) photodiodes with high sensitivity at 2 to 3 $\mu m$ wavelengths and a cut-off wavelength of 3.2 $\mu m$. The temperature of the optics is kept below 193 K (-80 deg C) to achieve the science requirement (less than 203 K at 1 km). The FOV is 0.11 deg x 0.11 deg. NIRS3 is equipped with two small halogen lamps to monitor the spectrometer performance in flight. The data taken with those lamps before and after the touchdown operations showed a performance decrease of ~20% due to dust contamination, but no other performance changes have been observed so far.

## 2.4. Laser Altimeter (LIDAR)

The light detection and ranging (LIDAR) laser altimeter consists of three components: a laser module, a telescope module that receives pulses, and a digital controller that measures the time of flight of the pulses (Mizuno et al., 2017). To cover a wider range (30 m – 25 km), the telescope module has a long-range system (Far system) and a short-range system (Near system). If the target is located within 1 km, the Near system is turned on to measure a short range. Otherwise, the Far system is operational. The Laser module uses an Yttrium Aluminum Garnet (YAG) laser with a frequency of 1,064 nm. The accuracy of measurement is $\pm 1$ m at 30 m altitude and $\pm 5.5$ m at 25 km altitude. The pre-flight test confirmed no degradation of the Laser Diode (LD) even after 16,320,000 LD shots [Mizuno et al., 2017]. LIDAR was designed to detect dust (Senshu et al., 2017) and measure albedo variations (Yamada et al., 2017).

Through the mission, the laser output has been observed to be stable. The telescope module has experienced dust contamination during the touchdown operations and a decrease of its sensitivity (around a few % for the NEAR system after the first touchdown and around 10% for the FAR system after the second touchdown). However, this effect does not significantly affect LIDAR's ranging measurement capabilities in the extended mission. As of November 2019, the laser module has experienced 7,077,853 shots, and we expect that LIDAR can still have ~9 million shots without major degradation.

## 3. Scientific motivations of extended mission
### 3.1. Formation and evolution processes of small bodies

After the sun's formation, small grains of dust and ice underwent accumulations in violent turbulence and became part of planetesimals in the solar nebular (Chiang and Youdin, 2010; Goldreich and Ward, 1973). The growth of planetesimals has been intensively debated and considered to happen in many different ways such as collisional agglomeration (Hartman and Davis, 1975; Weidenschilling and Cuzzi, 1993) and gravitational collapse (Birnstiel et al., 2016; Johansen et al., 2007), while a series of numerical and observational data analyses have suggested that planetesimals have once had sizes of tens to hundreds of km in diameter, or larger (Bottke et al., 2005; Delbo et al., 2017). Planetesimals' compositions strongly depend on the locations of accumulation. Over their lifetime, some planetesimals have been part of planets or larger small bodies, while others have been exposed to chaotic orbital perturbation and violent collisional processes. Asteroids are considered to be fossils of planetesimals (but heavily processed by numerous collision events). Numerical studies with empirical data fitting have shown that the collisional lifetime of asteroids having sizes of smaller than 10 – 100 km may be shorter than the age of the solar system, implying that such bodies have repeatedly experienced catastrophic disruption and reaccumulation (Bottke et al., 2005; Dohnanyi, 1969; Farinella et al., 1998).

The fragment size distributions are related to the collisional origin of small bodies. While depending on the kinetic energy input, fragments' size distributions correlate with their velocities and rotational distributions after catastrophic disruption; the smaller the size is, the faster the ejection velocity and rotation are (Giblin et al., 1998; Kadono et al., 2009). Such trends are also inferred from recently observed boulder distributions on asteroids (Michikami et al., 2019, 2010).

Violent collisional processes control the bulk structures of asteroids. There may be three structural types, which depend on the size (Walsh, 2018). First, asteroids larger than 10 km may possess fragmented layers on the top surface while having shattered blocks inside the body. Eros, observed by NEAR-Shoemaker, may possess such a structure, which was predicted by geomorphological observations (Cheng et al., 2002). Second, asteroids smaller than 10 km in diameter to 100 m are in general considered to be rubble piles, or gravitational aggregates of fragments. This idea is supported by the observed spin-size distribution (Figure 1) (Warner et al., 2009). In this size range, the spin period is in general bounded at ~2.3 h, which implies that if the bulk density is ~2,000 kg/m$^3$, rubble piles cannot spin faster than this limit as they may be disintegrated due to high centrifugal forces. Itokawa, Ryugu, and Bennu are considered to be rubble piles because of their highly porous conditions (Barnouin et al., 2019; Fujiwara et al., 2006; Watanabe et al., 2019). Third, an asteroid smaller than 100 m in diameter is in general monolithic, i.e., just a large boulder. This argument is also supported by observational evidence that asteroids at this size range can rotate at spin states higher than the gravitational spin limits (Figure 1). Such bodies may be the largest monolithic elements that may have escaped from being part of larger asteroids or planets when they were ejected at high speed, followed by catastrophic impact processes (Michel et al., 2001).

Shattered bodies like Eros (Cheng et al., 2002) and rubble pile bodies like Itokawa, Ryugu, and Bennu (Fujiwara et al., 2006; Lauretta et al., 2019a; Watanabe et al., 2019) were explored by rendezvous spaceflight missions. Also, reported are the physical properties of larger asteroids and their moons (if they exist) that were observed by flyby spaceflight missions (Belton et al., 1996; Coradini et al., 2011; Keller et al., 2010; Veverka et al., 1997; Yeomans et al., 1997). A key issue that remains unsolved is what asteroids smaller than 100 m in diameter look like. How do they evolve the orbital and geological conditions, given such a small size? Rendezvous operations with detailed remote sensing observations of these bodies will provide key information for the catastrophic disruption history of parent bodies. Furthermore, as they are rotating fast, they should resist against strong centrifugal forces. This severe condition will further constrain the evolution processes of these bodies.

## 3.2. Planetary defense

Planetary defense is a research field that attempts to monitor potentially hazardous objects and give proper assessment of effective strategies for deflecting them and has been paid attention recently in space engineering and science. There are several challenging issues to be solved. First, small asteroids are harder to be discovered and tracked by ground-based and space-based telescopes. While tireless observation efforts have been discovering new objects regularly, ~ 70% of Near-Earth Objects (NEOs) larger than 140 m in diameter have not yet been found, and the number of those less than that size is far from the predicted population (Harris and D'Abramo, 2015; Mainzer, 2017). This situation indicates that the majority of the small body population (especially extremely small ones) is unknown. Second, small asteroids still have enough kinetic energy to cause regional destruction and civil infrastructure (Huebner et al., 2009), as seen by the recent fire ball event in Chelyabinsk, Russia (Borovička et al., 2013; Brown et al., 2013). Similar fire ball events after the Chelyabinsk event were also observed (Borovička et al., 2017), meaning that such events frequently occur and can potentially threaten our population centers (Huebner et al., 2009). The mechanisms of fire ball events have been analyzed numerically, showing that the physical properties may control the shock wave propagations in the atmosphere from the ground zero (Aftosmis et al., 2019). Further studies are necessary to evaluate potential risks of physical, social, and environmental effects after asteroid impact events happen (Baum, 2018).

Growing efforts have proposed effective procedures for deflecting hazardous objects. Kinetic impactors and explosive processes have been one of common mitigation technologies considered in the communities (Leung et al., 2017; Wie et al., 2017), although the physical properties of target asteroids are key parameters of the deflection efficiency (Bruck Syal et al., 2016; Feldhacker et al., 2017). There have been proposed other deflection techniques using ablation (McMahon and Scheeres, 2017), mass ejection (Brack and McMahon, 2020), tethering (Venditti et al., 2020), and gravitational tractor (Lu and Love, 2005).

NASA's DART mission and ESA's Hera mission, jointly called the AIDA mission, define Planetary Defense as key part of the primary mission goals (Cheng et al., 2018, 2016; Michel et al., 2018, 2016) and are targeting the binary asteroid Didymos (Naidu et al., 2020). DART will send a spacecraft that impacts Dimorphos, the secondary of the Didymos system, in 2022 (Cheng et al., 2018), to assess the DART spacecraft's deflection capability while Hera will later conduct remote sensing observations after rendezvousing with it in 2026 (Michel et al., 2020b). Studies have shown that the deflection capability strongly depends on the environments and physical properties of the target asteroid, implying that detailed analyses are necessary to employ proper asteroid deflection technologies (Agrusa et al., 2020; Hirabayashi et al., 2019a, 2017; Raducan et al., 2020, 2019; Stickle et al., 2020).

Based on these arguments, there are three key issues.

1. Small bodies (especially smaller than 100 m in diameter) are the most common in the solar system and can influence our activities and civil infrastructure more frequently than larger bodies.
2. Observational capabilities have not yet reached high discovery completeness levels for small bodies, and only a small fraction of these bodies has been physically characterized.
3. Successful asteroid deflection strongly depends on the physical conditions of asteroids. Without the information, there would be large outcome uncertainties.

These issues cannot be solved without detailed proximity observations, and we address the necessity of proper spaceflight explorations to establish technological and scientific innovations for planetary defense.

## 4. Mission objectives of extended mission

The arguments in Section 3 directly connect with one common issue: *How can a body smaller than 100 m in diameter exist at a high spin rate?* Many asteroids of this size are rotating above the gravitational spin limit (Hergenrother and Whiteley, 2011). These bodies may simply be monolithic and leftovers that escaped from reaccumulation processes after parent bodies' catastrophic disruptions. Only rendezvous missions can detail their geological and geophysical conditions, which will provide key information for the issue. Addressing this question will give clear hints on (1) the formation and evolution processes of small bodies and (2) planetary defense. By the way, to achieve rendezvous operations, we also envision the mission's long-term cruise opportunity. This cruise period is key to advancing sophisticated technologies for achieving complex deep space explorations and scientific investigations that require long-term monitoring.

Importantly, observations of asteroids smaller than 100 m in diameter are not equivalent to those of similar-sized boulders on larger rubble pile asteroids. Large boulders on larger asteroids cannot be observed from multiple directions because spacecraft cannot easily approach them due to operational constraints and some part of the boulders are buried by regolith layers. Second, the rotation gives stronger constraints on their formation and evolution mechanisms.

Here, we define the following mission as the Hayabusa2 extended mission.

- The Hayabusa2 spacecraft will cruise for ~10 years until the early 2030s and conduct multiple swing-by and fly-by operations, followed by a rendezvous with a small object that rotates at fast spin.

To achieve this extended mission, we define three mission objectives (MOs). The items in each MO are labeled as either a scientific category (S) or an engineering category (E).

### *MO.1 Advances on long-term deep spaceflight operation technology and science*
- (S) Take advantage of a long-term cruise phase to monitor astrophysical phenomena.
- (S) Use flyby opportunities to conduct scientific investigations of flyby targets.

- (E) Demonstrate the capability of running the Japan-made Ion Engine System (IES) for a long-term period.
- (E) Establish Japan's multiple swing-by operation technologies for future deep spaceflight exploration missions.
- (E) Develop technologies for enhancing the operation efficiency to minimize necessary tasks for a long-term period.

### MO.2 Geological and geophysical characterizations of a small fast rotator.
- (S) Detail fast-rotating asteroids with sizes smaller than 100 m, objects that have poorly been observed from the ground- and space-based telescopes.
- (S) Investigate how they exist beyond their spin barrier to characterize their geological and geophysical conditions.
- (S) Constrain the formation and evolution of small bodies based on the observed physical conditions.

### MO.3 Technological and scientific developments and demonstrations for planetary defense.
- (S) Advance the scientific knowledge of the nature of small bodies that are common in the solar system and may threaten activities and civil infrastructure.
- (E) Establish technologies for accurately navigating, guiding, and controlling spacecraft towards a small asteroid.
- (SE) Develop innovative rendezvous proximity operations to significantly contribute to developing better strategies and planning for planetary defense.

MO.1 and MO.2 include key elements to rendezvous with the target object, as well as long-term observations during the cruise period. In MO.3, the second and third items contain technological innovations and enhance the technological innovations for small body proximity operations, which was also addressed in the nominal mission (Tsuda et al., 2019). The extended mission will promote Japanese space exploration technologies and scientific knowledge about planetary sciences, orchestrated with the Martian Moon Exploration (MMX) mission (Usui et al., 2020) and other planned missions led by JAXA and international partnerships. In the later discussions, we focus on the items labeled as (S) in each MO.

## 5. Selection of target asteroids
This section discusses how we selected the target candidates for rendezvous operations. We conducted a two-step selection process.

### 5.1. Step 1: Trajectory constraints
We first obtained the orbital information of small bodies on January 2019 from the JPL Small-Body Database. It included 18,002 small bodies (17,667 asteroids and 335 comets). By accounting for Earth and/or Venus gravity assists, we obtained trajectories for the spacecraft to reach the cataloged small bodies. The trajectory optimization process included preliminary assessments using the ballistic assumption and further analyses using low-thrust propulsion under the assumption that IES perfectly works (Specific Impulse, or Isp, of 2900 sec). The spacecraft's trajectory is constrained because of the capsule re-entry maneuvering around Earth at the end of December 2020. The initial state condition of the spacecraft for this search is given in Table 1.

We down selected target candidates by introducing six reachability criteria (RCs) (see Table 2). RC1 and RC2 define geometric constraints that eliminate complex trajectory searches, which could effectively save the assessment time to find accessible bodies. RC3 is a physical constraint on the spacecraft's maneuvering capability. We estimated that the spacecraft could still change its velocity, or ΔV, up to 1.6 km/sec. RC4 is rather arbitrary but limits the mission period to be less than 12 years (this can also constrain the total budget for the extended mission). RC5 constrains the target candidates with their orbital uncertainties. We used the condition code used in the JPL Small-Body Database, which is identical to the uncertainty parameter, $U$, defined by the Minor Planet Center (MPC) (footnote: https://www.minorplanetcenter.net/iau/info/UValue.html). $U$ is a function of the uncertainty in the perihelion time, the eccentricity, the orbital period, the uncertainty in the orbital period, and empirical data. It scales the uncertainty in orbit in the range between 0 and 9. 0 is the least uncertain, while 9 is the highest uncertain. We excluded all candidate bodies with $U > 5$. RC6 defines the distance from the sun to the spacecraft as being less than 1.5 AU, where IES is operational during the rendezvous phase. Using RC1 through RC6, we obtained 14 small bodies (no cometary bodies) as target candidates.

## 5.2. Step 2: Mission constraints

Using the selected 14 target candidates from Step 1, we further evaluated their constraints that include engineering and scientific values. We conducted a simple scaling analysis to grade each target candidate. We defined seven mission criteria (MCs), each of which was scaled from 0 to 10. In this analysis, 0 is the most preferred, while 10 is the least. The MCs are given in Table 3. Below, we describe each MC and its scaling process.

MC1 through MC4 define the physical properties of the target candidates. MC1 defines the size of the target candidates. Among the list of small bodies from Step 1, we found that 2015 TJ1 was the largest and might have a size of 107.3 m. We decided to use this size as 10 and linearly scale the size of a body, i.e., $(scale) = d/107.3 \times 10$, where $d$ is the diameter of a small body in meters. We removed any bodies with an absolute magnitude larger than 26. MC2 considers the spin periods of the target candidates. The longest spin period was chosen as 30 h to represent the spin periods of typical small bodies. Considering this spin period to be Scale 10, we introduced a scaling law, which is given as $(scale) = P/30 \times 10$, where $P$ is the spin period [h]. If there was no report for the spin period, the scale was set to be 10. Note that as there were only two out of the 14 candidates reported for their spin periods (1998 KY26 and 2001 AV43), this criterion became a strong constraint on the target candidates.

MC3 gives a scaling law for the shape. The detection of the shape can give significant information for planning of proximity operations prior to the arrival. In our scaling, we introduced the following conditions. First, we defined the scale as 0 if there exists a shape model driven by ground radar observations. Second, we considered the scale to be 3 if there was a rough estimate of the shape (especially the aspect ratio) driven by optical observations. Otherwise, we defined it as 10. The Database of Asteroid Models from Inversion Techniques (DAMIT) may offer shape models constructed by photometric observations (Ďurech et al., 2010). We found that there were no available shape models in this archive for our target candidates. It is in general challenging to constrain the shapes of asteroids. As we attempted to find smaller candidates, this issue was critical, and we found that almost all the asteroid candidates were barred by this constraint.

MC4 constrains the material composition, and we developed the following conditions. For Condition I, there is some information of material composition by spectroscopic observations. Condition II defines whether the target candidates are all but S-, C-, or B-type. This condition distinguishes the candidates from those observed by earlier rendezvous missions. While it would be scientifically beneficial to visit a target that exhibits the material compositions observed by earlier rendezvous missions, we decided to focus on the material compositions that have not been explored, yet. We excluded the taxonomic classes of Vesta (V-type) and Ceres (C, G-type) from Condition II; in other words, we accounted for V- and G-types as the material compositions that were not observed by earlier rendezvous missions because Vesta and Ceres are larger than our target sizes. If Conditions I and II were satisfied, we considered a candidate to be 0. If only Condition I was satisfied, then the target candidates were graded to be 5. Otherwise, the scale was 10.

MC5 evaluates whether the candidates can be observed by ground-based telescopes before the spacecraft arrival. If target candidates with $V < 20$, where $V$ is the apparent visible magnitude, can be observed by the 188-cm telescope at the Okayama Astrophysical Observatory, National Astronomical Observatory of Japan (NAOJ), we defined the scale as 0. If candidates with $V < 20$ are observable at the 8.2-meter telescope at Subaru, NAOJ, the scale was set to be 3. If candidates with $V < 25$ can be detected by the Subaru observations, the scale was 6. Otherwise, the scale was set to be 10. We note that although we used these telescopes as an example, the observability derived are applicable to other telescopes with similar capability.

MC6 describes the orbital uncertainties of the target candidates. Among the 14 target candidates, we would prefer one that has higher orbit determination accuracy so that ONC-T can properly find it during the approach phase. Because we have already used the $U$ code in Step 1, all the candidates in Step 2 should have $U \leq 5$. We simply multiplied $U$ by a factor of 2 to define the $0 - 10$ scale for this criterion.

MC7 defines the existence of backup targets. This criterion guarantees the achievement of a rendezvous exploration mission by having backup targets in case we find difficulty in arriving at the primary target. We defined the scale as 0 if there are backup targets on the course of the planned trajectory. Otherwise, the scale was set to be 10.

Given all these seven criteria, we obtained the total score that ranges from 0 (best) to 70 (least). Table 4 shows the scores of the 14 target candidates for rendezvous operations from Step 1. We find that 1998 KY26 and 2001 AV43 are the most favorable objects for the extended mission target candidates.

## 6. Mission scenarios

Given our two favorable target candidates, we propose the rendezvous scenario for each candidate. For the scenario of a rendezvous with 2001 AV43, the trajectory will need one Venus gravity assist and two Earth gravity assists after returning to Earth in 2020 (the EVEEA scenario). On the other hand, for that of a rendezvous with 1998 KY26, it will require two Earth gravity assists after 2020 (the E(A)EEA scenario). If the feasibility is confirmed, the spacecraft may be able to fly by asteroid 2001 CC21. Figure 2 illustrates key spacecraft operations in both scenarios.

## 6.1. EVEEA scenario

The final goal is to rendezvous with 2001 AV43. After the spacecraft departs from Earth in December 2020, it will be in cruise operation until 2024. In mid-2024, the spacecraft will perform a Venus swing-by. In 2025 and 2026, the spacecraft will perform two Earth swing-bys. Finally, in late 2029, it will arrive at 2001 AV43. Under the assumption that Isp = 2,900 sec for the spacecraft's IES, the planned trajectory will need additional ΔV of 1.45 km/sec. The distance from the sun is around 1 AU when the spacecraft arrives at this asteroid. Figure 3 and Table 5 show the planned mission scenario and trajectory, respectively.

## 6.2. E(A)EEA scenario

The spacecraft is planned to arrive at 1998 KY26 in late 2031. After the departure from Earth in December 2020, the spacecraft will continue its cruise operation. In middle 2026, the spacecraft may fly by 2001 CC21. It will approach the asteroid with high speed (~ 5 km/sec) during the flyby, and thus sophisticated guidance, navigation, and control technologies are necessary. Note that the spacecraft was not designed to conduct high-speed flybys at small bodies. Further assessments will thus solidify the flyby operation with limited capabilities. After two Earth swing-bys in late 2027 and early 2028, the spacecraft will arrive at the target asteroid. The planned trajectory will need additional ΔV of 1.09 km/sec, given the 2,900-sec Isp. The distance from the sun is ~1 AU when the spacecraft arrives at this asteroid. The planned operation schedule and trajectory are given in Table 6 and Figure 4, respectively.

## 7. Rendezvous with small asteroids
### 7.1. 2001 AV43

2001 AV43 is the target candidate for the EVEEA scenario. Table 7 shows its physical properties. This asteroid is reported to have a diameter of about 40 m and a spin period of 10.2 min (Hergenrother and Whiteley, 2011). No obvious tumbling motion was reported (Hergenrother and Whiteley, 2011). The shape of this asteroid may be elongated, and the aspect ratio (the short axis/the long axis) may be around 0.5, according to light curve observations (Hergenrother and Whiteley, 2011). The taxonomic class is uncertain although this asteroid exhibits a phase coefficient suggesting a similar albedo to S-type asteroids. This asteroid is listed online (footnote: https://cneos.jpl.nasa.gov/nhats/) and considered to be accessible by NASA's human spaceflight missions.

We computed the surface slope distribution (Figure 5) and the stress distribution (Figure 6) by considering the estimated physical properties (Table 7). As the shape is not well known, we assumed that it is a biaxial ellipsoid. Also, we defined the bulk density as 2000 kg/m$^3$ based on the physical properties of Itokawa, an S-type asteroid (Fujiwara et al., 2006). Figure 5 describes 2001 AV43's high slope distribution. The spin axis is along the z axis. In the equatorial regions, the slope reaches 180 deg, which gives outward acceleration. Particles in these regions should be lofted unless there is attraction from the surface such as cohesion. Also, asteroids are often exposed to many physical processes including meteoroid impacts (Hirata et al., 2020; Lauretta et al., 2019a, 2019b; Sugita et al., 2019; Walsh et al., 2019), thermal cracking (Molaro et al., 2020b, 2020a), or electrostatic levitation (Hartzell, 2019). When the surface of 2001 AV43 is exposed to these processes, the removal of surface materials may have occurred continuously. Because of its outward acceleration, the material removal may gradually erode the surface morphology (Nagao et al., 2011).

Figure 6 describes the distribution of the minimum cohesive strength, $Y^*$, that can avoid structural failure due to severe loadings. If the actual cohesive strength is lower than $Y^*$, a given element should fail structurally (Hirabayashi et al., 2020; Nakano and Hirabayashi, 2020). Because this asteroid is elongated, and the spin period is short, a higher stress tends to appear in regions around the spin axis. $Y^*$ reaches 20 Pa at the center of the body. Regions far from the spin axis become less sensitive to structural failure; $Y^*$ becomes down to a few pascals at the edges of the body. Using observational data, reports have shown that rubble pile asteroids may have a cohesive strength of ~100 Pa (Hirabayashi et al., 2019b, 2014; Hirabayashi and Scheeres, 2015; Rozitis et al., 2014; Watanabe et al., 2019). These findings were consistent with the results from studies that analyzed van der Waals forces (Hartzell and Scheeres, 2011; Scheeres et al., 2010).

As seen from Figures 5 and 6, the 10.2-min spin period of this asteroid causes outward acceleration over the surface and tension inside the body, implying that cohesion is necessary to keep the present shape. A favorable explanation of this condition is that this asteroid is monolithic (Hergenrother and Whiteley, 2011). On the other hand, given the magnitude of cohesion, ~20 Pa, there is another possibility that rubble piles with tens of meters in diameter can also resist strong centrifugal force to keep their structural conditions (Hartzell and Scheeres, 2011; Holsapple, 2007; Scheeres et al., 2010).

The short spin state may have resulted from many different processes such as random spinning processes during mass ejection after catastrophic processes, micrometeoroid impacts, tidal torques, or solar radiation pressure. Because of their small sizes, their moments of inertia are small, leading to dramatic changes in rotation in a short timescale. If the structure is rubble pile, there is another constraint that the body has avoided reaccumulation to larger bodies and later formed its rubble pile structure when the spin period was long; otherwise, fragmented particles cannot gather gravitationally. YORP, micrometeoroid impacts, and tidal torques would have changed its spin period rapidly. A key issue of this process is whether rubble piles can form at this size, given extremely small gravity. Therefore, the monolithic structure would still be a favorable scenario to explain the structural conditions of 2001 AV43. This issue is a critical question to be answered by the extended mission, constraining a lower limit of the rubble pile size.

2001 AV43 has an observation arc of 13 years that includes radar astrometry. Thus, orbital uncertainties are small and enable the spacecraft to arrive at this asteroid with proper guidance, navigation, and control. Prior to the planned arrival at this asteroid in 2029, this asteroid is observable in 2023, 2026, and 2029. In 2029, the spacecraft may be able to arrive at the asteroid before the closest approach of 2001 AV43 to Earth. On November 11, 2019, 2001 AV43 reaches $3.1 \times 10^5$ km from the geocenter, which is within a lunar distance. This opportunity will enable collaboration explorations of this asteroid among the extended mission and ground based observation campaigns.

## 7.2. 1998 KY26

1998 KY26 is the primary target in the E(A)EEA scenario. This asteroid has been observed by optical observations and ground radar in detail (Ostro et al., 1999). The shape (Figure 7) was constrained based on the Doppler-based model, and thus there may be some uncertainties (Ostro et al., 1999). The shape model shows no prominent asymmetric features of this body. The physical properties are given in Table 8. The current estimate of the spin period is 10.7 min, and the size is about 20 m – 40 m. The surface color is reported to be dark, implying the potential taxonomic classes associated with mixture of carbonaceous materials and mafic silicates (B, C, F, G, D, P) (Ostro et al., 1999). The observations inferred a surface bulk density of 2,800 kg/m³ and roughness at centimeter-to-decimeter scales, suggesting that the surface may be exposed bear rocks (Ostro et al., 1999). If the material compositions are indeed carbonaceous and similar to Murchison's (Wopenka et al., 2013), the bulk density may be lower than the reported density. This asteroid is also listed online (footnote: https://cneos.jpl.nasa.gov/nhats/) and considered accessible by NASA's human spaceflight missions.

Figure 7 describes the surface slope of this asteroid. The equatorial region is mainly influenced by the centrifugal force, and the surface slope reaches 180 deg. Particles on the surface can be lofted if there is no attraction. Loose materials may not exist in these regions unless there is cohesion. If such lofting processes are common, there may exist dust particles orbiting around the body. Figure 8 characterizes the minimum cohesive strength distribution over the slice of the body (the z axis is defined along the spin axis). The internal region requires a cohesive strength of ~20 Pa. The structural condition is similar to that of 2001 AV43 around the central region. However, because of the round shape, the contours of the cohesive strength are rounder than those in 2001 AV43. If there is not enough strength to keep the current shape, the shape fails structurally and is fragmented into multiple large pieces (Hirabayashi and Scheeres, 2019). However, the derived cohesive strength is ~20 Pa and not as high as that of rocks (~ from MPa to GPa). While there are many conditions that would be favored for monolithic bodies, the rubble pile structure of 1998 KY26 is still possible.

Two additional astrometric observations during a second apparition in 2002 improved 1998 KY26's orbital uncertainty to ~ 5,000 km (1-$\sigma$) in July 2031, which is small enough for the spacecraft to detect this asteroid during the approach phase. However, further observations will reduce the uncertainty and give critical information of the physical properties. There are opportunities for ground-based observations in 2020, 2024, and 2028. In 2024, the V-magnitude will reach down to ~20. Observations during these opportunities will reduce the uncertainty down to ~ $2 \times 10^2$ km (1-$\sigma$).

## 7.3. Scientific constraints on proximity operations around the target asteroids

This section introduces the potential proximity operations. Because 2001 AV43 and 1998 KY26 have similar sizes, the operation processes will not be different. During the proximity operations at Ryugu, the home position (HP), where the spacecraft could stand by with zero-velocity, was defined ~20 km from the body along the Earth-asteroid direction. The Hayabusa2 team used remote sensing data taken at HP to assess descent operations. ONC-T achieved capturing Ryugu with ~500 pix and a pixel resolution of 2 m. If the same resolution is required from HP for the present target candidates, HP should be 750 m from the surface, and from this location, the resolutions of images from ONC-T and TIR will be 0.08 m/pix and 0.7 m/pix, respectively. NIRS3 will achieve a resolution of 1.5 m.

However, the constraints on HP (and even whether we define HP for the extended mission) strongly depend on engineering conditions such as the determination of the spacecraft's relative orbit to the target object (Tsuda et al., 2020) and descent operations (Ono et al., 2020). For Ryugu, a hybrid approach that combined radiometric-optical navigation and a stochastic-constrained optimum guidance method achieved reducing a relative position uncertainty of less than 100 m and a relative velocity uncertainty of less than 1 cm/s (Tsuda et al., 2020). The nominal mission conducted descent operations by using a ground control point navigation (GCP-NAV), in which guidance, navigation, and control (GNC) processes were conducted based on on-board and on-ground guidance systems and image-based navigation techniques (Ono et al., 2020). However, because the operation conditions are different, new engineering strategies need to be assessed. In this report, we do not outline proximity operations but argue necessary investigations to achieve the scientific investigations in MO.2 and MO.3. Figure 9 illustrates the resolution and maximum pixel size as a function of the distance from the body surface. Note that the following discussions are subject to change based on future scientific and engineering assessments.

We propose four science objectives (SOs) that address MO.2 and MO.3 (Table 9). SO.1 defines detailed observations of the morphological conditions and addresses MO.2 and MO.3. On Ryugu and Bennu, particle sizes of a few cm are a major part of the particle size distributions, and boulder crack sizes range down to less than 1 m (Molaro et al., 2020b; Sugita et al., 2019; Walsh et al., 2019). To observe such trends, we consider the required image resolution of ONC-T to be 1 cm. This requirement leads to a spacecraft altitude requirement of 90 m from the surface. Because of the target candidate's short spin period, ~10 min, a short stay longer than the spin period at this altitude will enable mapping surface regions from all the directions. However, the asteroid spin orientations may constrain global mapping conditions, which needs further assessment.

SO.2 characterizes the distributions of the spectral signature due to material heterogeneity and addresses MO.2 and MO.3. Applying NIRS3 and ONC-T, we will map the surface material conditions in the visible-near infrared wavelength range (seven filters between 0.39 $\mu m$ and 0.95 $\mu m$ for ONC-T and 1.8 − 3.2 $\mu m$ for NIRS3). NIRS3's original design can directly be applicable to carbonaceous asteroids. While not well identified in detail yet, 1998 KY26 exhibits its dark surface, implying a possibility of the carbonaceous nature. If this is the case, NIRS3 can directly characterize the details of surface compositions including the existence of hydrated minerals on this asteroid. For 2001 AV43, a potential S-type asteroid, we anticipate major absorption at 1 $\mu$m and 2 $\mu$m, which indicates the presence of olivine and pyroxene (Binzel et al., 2010). A combination of NIRS3's and ONC-T's observations will allow for characterizing such surface material compositions in detail. One observational constraint on ONC-T is that while it needs to rotate the filter wheel and set a proper filter, this process takes time. As the target candidate rotates at a short spin period, it may be challenging to take images with multiple filters at a given location during one asteroid rotation. To solve this issue, we plan to limit the use of filters or wait for the target location to come back after one rotation.

SO.3 explores the environment surrounding the target object to address MO.2. We will explore whether there are ejected particles. If small particles exist on the surface and depart from there due to physical processes, they should escape from the body because the surface velocity around the equator is already higher than the particle escape velocity. Combining ONC-T, TIR, and LIDAR, we will explore whether dust particles exist around the target body. LIDAR was designed to detect dust clouds surrounding the target body (Senshu et al., 2017). If the spacecraft is at an altitude of 1 km from the target, the detection limit of the number density may be $10^5$ m$^{-3}$ (Senshu et al., 2017).

SO.4 addresses MO.3 by conducting thermal analyses and identifying the surface materials. It was reported that Ryugu's materials were highly porous and structurally weak (Okada et al., 2020). This finding implies that among many meteorites falling onto the Earth ground, weakly structured ones were burned out in the atmosphere (Okada et al., 2020). Thus, airburst events may occur more often than impact cratering processes on the ground, although they are highly controlled by the structure of an asteroid (Brown et al., 2013). We will analyze whether this asteroid is such a weak object, using TIR to observe the global distribution of the brightness temperature and the thermal inertia. The distance from the sun to the target (for both 1998 KY26 and 2001 AV43) is ~ 1 AU during the approach phase, which is comparable with that of Ryugu during the arrival of Hayabusa2 spacecraft. Thus, TIR is expected to measure the surface temperature within well-calibrated temperature range (< 400 K). However, as the target is a fast rotator, the daytime temperature variation becomes flatter than Ryugu's. It is our future work to analyze how to constrain the thermal inertia given the flat temperature variation and the surface roughness. Also, observations of the YORP and Yarkovsky effects will help constrain the thermal and physical properties of the target. Models have predicted 1998 KY26's spin rate acceleration as $\sim(1-4) \times 10^{-7}$ s$^{-1}$yr$^{-1}$ (Nesvorný and Vokrouhlický, 2008, 2007). Long-term monitoring will play a crucial role in determining these effects and supporting the proximity observations.

We finally note that a touchdown will be considered as a possible operation. During the nominal mission, using target markers significantly helped the touchdown operations, which successfully allowed the spacecraft to identify the touchdown locations (Kikuchi et al., 2020; Ogawa et al., 2020). Although the spacecraft has one target marker left, the target body is rotating at a short spin period, and the majority of the surface experiences outward acceleration, preventing the extended mission from using it. The potential solution may be to track the natural features actively without deploying the target marker.

## 8. Scientific investigations during long-term cruise phases and flybys

### 8.1. Flyby observation of Venus

If the EVEEA scenario is preferred and selected, the spacecraft will fly by Venus in 2024. Venus has thick atmosphere, and its rotation period is 243 days. Because of this slow rotation, the atmosphere is intensely heated by the sun, giving complex energy and momentum exchanges over the entire atmosphere (generated atmospheric waves are called thermal tides). The global rotation of the atmosphere is much faster than that of the surface, so-called super-rotation, which is driven by complex systems of vertical and horizontal angular momentum transfer (Ando et al., 2018; Horinouchi et al., 2020, 2017; Kouyama et al., 2019, 2012). Moreover, the atmosphere in upper layers may also be affected by the circulation in lower layers that are controlled by local topography (Fukuhara et al., 2017). Venus' atmosphere has been closely monitored from 2006 to 2020 over 15 years. This long-term observation revealed that Venus has a 10-year-scale change in the UV albedo of the cloud at global scale and thus the strength of the super-rotation (Kouyama et al., 2013; Lee et al., 2019). This finding indicates that Venus changes its weather pattern due to the energy input from the Sun. However, this long-term observation was incomplete to characterize the long-term weather pattern, missing a critical piece of understanding its periodic variation.

The 2024 Venus flyby will be a critical timing to observe Venus' atmosphere. The Akatsuki mission is supposed to be completed before the spacecraft flyby. Therefore, combinations of data from Venus Express, Akatsuki, and Hayabusa2 will strongly constrain the periodic variation in the atmosphere. Using ONC, we will detail the albedo change. Using TIR will also allow for observing the temperature of Venus atmosphere. Given the planned trajectory, the most preferable observation chance will be $10-20$ days before the closest approach. In addition, recent studies have shown that the orbits of many exoplanets are similar to the orbital condition of Venus, implying that Venus is a planet whose conditions are common in exoplanets. Therefore, detailed observations of Venus can link the atmospheric conditions of exoplanets.

### 8.2. Flyby observations of 2001 CC21

2001 CC21 is the flyby target in the E(A)EEA scenario. If the E(A)EEA scenario is selected, the spacecraft may approach this asteroid at an approach speed of ~5 km/sec (the operation feasibility is currently assessed). This body's physical properties are listed in Table 10 (Binzel et al., 2004; Warner et al., 2009). The information is available online (footnote:http://www.as.utexas.edu/astronomy/research/people/ries/pha_photometry.html). The size is ~700 m, and the spin period is ~5 h. While the taxonomic class is not well identified, an L-type may be possible.

If the flyby is feasible, the spacecraft may reach 100 km from the target body during the closest approach. This will allow ONC-T to take images of this body at a resolution of ~10 m (about 70 pix) to observe the global shape and the distributions of large craters and boulders (larger than 30 - 40 m in size). TIR can also capture this object at a resolution of ~90 m (~8 pix), which may be high enough to observe the thermal heterogeneities on a global scale. The 5 km/sec approach speed may constrain the observation feasibility. It may be challenging to rotate the ONC filters because it takes a second to switch them, limiting the imaging opportunities. The potential solution may be to use a single filter to focus on the surface morphology without color variations.

## 8.3. Zodiacal light

Zodiacal light is a phenomenon in which there is a faint glow in the night sky. The source of the zodiacal light is a scattering of the sunlight due to the existence of interplanetary dust, which would be the smallest solid element that exists in the solar system. The distribution of interplanetary dust is considered to be proportional to $1/R$, given the assumption that the Poynting-Robertson drag is a main contributor to the dust distribution. However, the spatial distribution is still not well understood. Better understandings of the interplanetary dust distribution will characterize the debris disk in the solar system (Poppe et al., 2019), as well as giving stronger constraints on the cosmic optical background (Zemcov et al., 2017). Using ONC-T, we plan to attempt long-term observations of the zodiacal light. By monitoring the intensity at given sun-spacecraft phase angles and distances, we will constrain the spatial distribution. We plan to use the v-band and the wide band to observe the zodiacal light. The observations will be performed with ONC-T's maximum integration time of ~178 sec. Because the zodiacal light is expected to be dim, stray light may challenge the observations (Suzuki et al., 2018). We will solve this issue by controlling the spacecraft attitude so that ONC-T can avoid the stray light effect and properly face to observe the zodiacal light. We plan to separate the long-term cruise phase (~10 years) into multiple phases to effectively maximize observational opportunities.

## 8.4. Exoplanet search

Numerous exoplanets have been found recently by space telescopes including Kepler and TESS, which observe the transits of exoplanets with respect to their host stars. Although a great number of exoplanets have been found by transit photometry, the orbital period (i.e., transit timing) of many of them have not been determined very accurately. In particular, the transit timings of exoplanets around bright stars are particularly important to determine because they are valuable for spectroscopic observations to characterize exoplanetary atmospheres in the future. Bright stars do not require large aperture, such as NASA's Kepler and even TESS satellites. Light flux calculation shows that the small effective aperture (15.1 mm) of ONC-T can achieve 0.1% of shot noise for $8^{th}$ magnitude of stars. Because there are many other noise sources, actual error will be greater, but the conventional aperture photometry error for light curve of Ryugu ($9.6^{th}$ mag) from 1.5 million km away was <1% (Tatsumi et al., 2019). More advanced methods such as the pixel-level decorrelation (PLD) method (Deming et al., 2015) would substantially help reduce the error closer to the shot noise. We will test the capability of detecting the actual transits of observed exoplanets. The potential targets include KELT-11b (Pepper et al., 2017) and KELT-19Ab (Siverd et al., 2017), both of which have a few hours of transit time. We will use ONC-T to take images every 5 min about twice the length of transit time to cover the baseline light curve, leading to ~100 images in total. Once we validate the proposed approach, we will expand observation targets to exoplanets with needs for accurate transit measurements.

## 9. Conclusion

This study proposed scientific investigations in the Hayabusa2 extended mission that attempts to rendezvous with a small asteroid rotating beyond its spin limit by using the Hayabusa2 spacecraft after its nominal mission is completed in December 2020. We discussed two mission scenarios. In the first scenario (EVEEA), the spacecraft would perform swing-by maneuvers at Venus once and the Earth twice to approach 2001 AV43, a 40-m-diameter asteroid rotating at a spin period of 10.2 min, in 2029. In the second scenario (E(A)EEA), after two Earth swing-bys, the spacecraft would arrive at 1998 KY26, a 30-m-diameter asteroid spinning at a spin period of 10.7 min, in 2031. This scenario would also include a possible flyby of 2001 CC21 in 2026. Both scenarios would have a ~10-year-long cruise phase and take advantage of this period to monitor physical phenomena in space. Given the proposed operations, the Hayabusa2 extended mission will significantly contribute to the scientific understanding of the formation and evolution processes of small bodies, which contributes to addressing the mechanisms of material transport in the solar system, and planetary defense.

## 10. Acknowledgments

The authors greatly thank Mike Nolan, Marco Micheli, Bill Ryan, and Tim Lister for constructive discussions that helped our scientific assessments regarding the observation conditions and orbit uncertainties of the target candidates. Part of this research was carried out at the Jet Propulsion Laboratory, California Institute of Technology, under a contract with NASA.

**Reference:**

Aftosmis, M.J., Mathias, D.L., Tarano, A.M., 2019. Simulation-based height of burst map for asteroid airburst damage prediction. Acta Astronaut. 156, 278–283.

Agrusa, H.F., Richardson, D.C., Davis, A.B., Fahnestock, E., Hirabayashi, M., Chabot, N.L., Cheng, A.F., Rivkin, A.S., Michel, P., 2020. A benchmarking and sensitivity study of the full two-body gravitational dynamics of the DART mission target, binary asteroid 65803 Didymos. Icarus 349.

Ando, H., Takagi, M., Fukuhara, T., Imamura, T., Sugimoto, N., Sagawa, H., Noguchi, K., Tellmann, S., Pätzold, M., Häusler, B., Murata, Y., Takeuchi, H., Yamazaki, A., Toda, T., Tomiki, A., Choudhary, R., Kumar, K., Ramkumar, G., Antonita, M., 2018. Local Time Dependence of the Thermal Structure in the Venusian Equatorial Upper Atmosphere: Comparison of Akatsuki Radio Occultation Measurements and GCM Results. J. Geophys. Res. Planets 123, 2270–2280.

Arai, T., Nakamura, T., Tanaka, S., Demura, H., Ogawa, Y., Sakatani, N., Horikawa, Y., Senshu, H., Fukuhara, T., Okada, T., 2017. Thermal Imaging Performance of TIR Onboard the Hayabusa2 Spacecraft. Space Sci. Rev. 208, 239–254.

Arakawa, M., Saiki, T., Wada, K., Ogawa, K., Kadono, T., Shirai, K., Sawada, H., Ishibashi, K., Honda, R., Sakatani, N., Iijima, Y., Okamoto, C., Yano, H., Takagi, Y., Hayakawa, M., Michel, P., Jutzi, M., Shimaki, Y., Kimura, S., Mimasu, Y., Toda, T., Imamura, H., Nakazawa, S., Hayakawa, H., Sugita, S., Morota, T., Kameda, S., Tatsumi, E., Cho, Y., Yoshioka, K., Yokota, Y., Matsuoka, M., Yamada, M., Kouyama, T., Honda, C., Tsuda, Y., Watanabe, S., Yoshikawa, M., Tanaka, S., Terui, F., Kikuchi, S., Yamaguchi, T., Ogawa, N., Ono, G., Yoshikawa, K., Takahashi, T., Takei, Y., Fujii, A., Takeuchi, H., Yamamoto, Y., Okada, T., Hirose, C., Hosoda, S., Mori, O., Shimada, T., Soldini, S., Tsukizaki, R., Iwata, T., Ozaki, M., Abe, M., Namiki, N., Kitazato, K., Tachibana, S., Ikeda, H., Hirata, N., Hirata, N., Noguchi, R., Miura, A., 2020. An artificial impact on the asteroid (162173) Ryugu formed a crater in the gravity-dominated regime. Science (80-. ). 368, 1–5.

Arakawa, M., Wada, K., Saiki, T., Kadono, T., Takagi, Y., Shirai, K., Okamoto, C., Yano, H., Hayakawa, M., Nakazawa, S., Hirata, N., Kobayashi, M., Michel, P., Jutzi, M., Imamura, H., Ogawa, K., Sakatani, N., Iijima, Y., Honda, R., Ishibashi, K., Hayakawa, H., Sawada, H., 2017. Scientific Objectives of Small Carry-on Impactor (SCI) and Deployable Camera 3 Digital (DCAM3-D): Observation of an Ejecta Curtain and a Crater Formed on the Surface of Ryugu by an Artificial High-Velocity Impact. Space Sci. Rev. 208, 187–212.

Barnouin, O.S., Daly, M.G., Palmer, E.E., Gaskell, R.W., Weirich, J.R., Johnson, C.L., Asad, M.M.A., 2019. SUPPLEMENTARY INFORMATION Shape of ( 101955 ) Bennu indicative of a rubble pile with internal stiffness. Nat. Geosci. 12, 247–252.

Barucci, M.A., Hasselmann, P.H., Fulchignoni, M., Honda, R., Yokota, Y., Sugita, S., Kitazato, K., Deshapriya, J.D.P., Perna, D., Tatsumi, E., Domingue, D., Morota, T., Kameda, S., Iwata, T., Abe, M., Ohtake, M., Matsuura, S., Matsuoka, M., Hiroi, T., Nakamura, T., Kouyama, T., Suzuki, H., Yamada, M., Sakatani, N., Honda, C., Ogawa, K., Hayakawa, M., Yoshioka, K., Cho, Y., Sawada, H., Takir, D., Vilas, F., Hirata, N., Hirata, N., Tanaka, S., Yamamoto, Y., Yoshikawa, M., Watanabe, S.,

Tsuda, Y., 2019. Multivariable statistical analysis of spectrophotometry and spectra of (162173) Ryugu as observed by JAXA Hayabusa2 mission. Astron. Astrophys. 629, 1–10.

Baum, S.D., 2018. Uncertain human consequences in asteroid risk analysis and the global catastrophe threshold. Nat. Hazards 94, 759–775.

Belton, M.J.S., Chapman, C.R., Klaasen, K.P., Harch, A.P., Others, 1996. Galileo's Encounter with 243 Ida: An Overview of the Imaging Experimen. Icarus 120, 1–19.

Bibring, J.P., Hamm, V., Langevin, Y., Pilorget, C., Arondel, A., Bouzit, M., Chaigneau, M., Crane, B., Darié, A., Evesque, C., Hansotte, J., Gardien, V., Gonnod, L., Leclech, J.C., Meslier, L., Redon, T., Tamiatto, C., Tosti, S., Thoores, N., 2017. The MicrOmega Investigation Onboard Hayabusa2. Space Sci. Rev. 208, 401–412.

Binzel, R.P., Morbidelli, A., Merouane, S., Demeo, F.E., Birlan, M., Vernazza, P., Thomas, C.A., Rivkin, A.S., Bus, S.J., Tokunaga, A.T., 2010. Earth encounters as the origin of fresh surfaces on near-Earth asteroids. Nature 463, 331–334.

Binzel, R.P., Perozzi, E., Rivkin, A.S., Rossi, A., Harris, A.W., Bus, S.J., Valsecchi, G.B., Slivan, S.M., 2004. Dynamical and compositional assessment of near-Earth object mission targets. Meteorit. Planet. Sci. 39, 351–366.

Birnstiel, T., Fang, M., Johansen, A., 2016. Dust Evolution and the Formation of Planetesimals. Space Sci. Rev. 205, 41–75.

Borovička, J., Spurný, P., Brown, P., Wiegert, P., Kalenda, P., Clark, D., Shrbený, L., 2013. The trajectory, structure and origin of the Chelyabinsk asteroidal impactor. Nature 503, 235–237.

Borovička, J., Spurný, P., Grigore, V.I., Svoreň, J., 2017. The January 7, 2015, superbolide over Romania and structural diversity of meter-sized asteroids. Planet. Space Sci. 143, 147–158.

Bottke, W.F., Durda, D.D., Nesvorný, D., Jedicke, R., Morbidelli, A., Vokrouhlický, D., Levison, H.F., 2005. Linking the collisional history of the main asteroid belt to its dynamical excitation and depletion. Icarus 179, 63–94.

Brack, D.N., McMahon, J.W., 2020. Active mass ejection for asteroid manipulation and deflection. J. Spacecr. Rockets 57, 413–433.

Brown, P.G., Assink, J.D., Astiz, L., Blaauw, R., Boslough, M.B., Borovička, J., Brachet, N., Brown, D., Campbell-Brown, M., Ceranna, L., Cooke, W., De Groot-Hedlin, C., Drob, D.P., Edwards, W., Evers, L.G., Garces, M., Gill, J., Hedlin, M., Kingery, A., Laske, G., Le Pichon, A., Mialle, P., Moser, D.E., Saffer, A., Silber, E., Smets, P., Spalding, R.E., Spurný, P., Tagliaferri, E., Uren, D., Weryk, R.J., Whitaker, R., Krzeminski, Z., 2013. A 500-kiloton airburst over Chelyabinsk and an enhanced hazard from small impactors. Nature 503, 238–241.

Bruck Syal, M., Michael Owen, J., Miller, P.L., 2016. Deflection by kinetic impact: Sensitivity to asteroid properties. Icarus 269, 50–61.

Cheng, A.F., Izenberg, N., Chapman, C.R., Zuber, M.T., 2002. Ponded deposits on asteroid 433 Eros. Meteorit. Planet. Sci. 37, 1095–1105.

Cheng, A.F., Michel, P., Jutzi, M., Rivkin, A.S., Stickle, A., Barnouin, O., Ernst, C., Atchison, J., Pravec, P., Richardson, D.C., 2016. Asteroid Impact & Deflection Assessment mission: Kinetic impactor. Planet. Space Sci. 121, 27–35.

Cheng, A.F., Rivkin, A.S., Michel, P., Atchison, J., Barnouin, O., Benner, L., Chabot, N.L., Ernst, C., Fahnestock, E.G., Kueppers, M., Pravec, P., Rainey, E., Richardson, D.C., Stickle, A.M., Thomas, C., 2018. AIDA DART asteroid deflection test: Planetary defense and science objectives. Planet. Space Sci. 157, 104–115.

Chiang, E., Youdin, A.N., 2010. Forming Planetesimals in Solar and Extrasolar Nebulae, Annual Review of Earth and Planetary Sciences.

Coradini, A., Capaccioni, F., Erard, S., Arnold, G., De Sanctis, M.C., Filacchione, G., Tosi, F., Barucci, M.A., Capria, M.T., Ammannito, E., Grassi, D., Piccioni, G., Giuppi, S., Bellucci, G., Benkhoff, J., Bibring, J.P., Blanco, A., Blecka, M., Bockelee-Morvan, D., Carraro, F., Carlson, R., Carsenty, U., Cerroni, P., Colangeli, L., Combes, M., Combi, M., Crovisier, J., Drossart, P., Encrenaz, E.T., Federico, C., Fink, U., Fonti, S., Giacomini, L., Ip, W.H., Jaumann, R., Kuehrt, E., Langevin, Y., Magni, G., McCord, T., Mennella, V., Mottola, S., Neukum, G., Orofino, V., Palumbo, P., Schade, U., Schmitt, B., Taylor, F., Tiphene, D., Tozzi, G., 2011. The surface composition and temperature of asteroid 21 lutetia as observed by Rosetta/VIRTIS. Science (80-. ). 334, 492–494.

Delbo, M., Walsh, K., Bolin, B., Avdellidou, C., Morbidelli, A., 2017. Identification of a primordial asteroid family constrains the original planetesimal population. Science (80-. ). 357, 1026–1029.

Deming, D., Knutson, H., Kammer, J., Fulton, B.J., Ingalls, J., Carey, S., Burrows, A., Fortney, J.J., Todorov, K., Agol, E., Cowan, N., Desert, J.M., Fraine, J., Langton, J., Morley, C., Showman, A.P., 2015. Spitzer secondary eclipses of the dense, modestly-irradiated, giant exoplanet hat-P-20b using pixel-level decorrelation. Astrophys. J. 805, 132.

Dohnanyi, J.S., 1969. Collisional model of asteroids and their debris. J. Geophys. Res. 74, 2531–2554.

Ďurech, J., Sidorin, V., Kaasalainen, M., 2010. Astrophysics DAMIT : a database of asteroid models. Astron. Astrophys. A46, 13.

Farinella, P., Vokrouhlicky, D., Hartmann, W.K., 1998. Meteorite delivery via Yarkovsky orbital drift (vol 132, pg 378, 1998). Icarus 134, 347.

Feldhacker, J.D., Bruck Syal, M., Jones, B.A., Doostan, A., McMahon, J., Scheeres, D.J., 2017. Shape dependence of the kinetic deflection of asteroids. J. Guid. Control. Dyn. 40, 2417–2431.

Fujiwara, A., Kawaguchi, J., Yeomans, D.K., Abe, M., Mukai, T., Okada, T., Saito, J., Yano, H., Yoshikawa, M., Scheeres, D.J., Cheng, A.F., Demura, H., Gaskell, R.W., Hirata, N., Ikeda, H., Kominato, T., Miyamoto, H., Nakamura, A.M., Nakamura, R., Sasaki, S., Uesugi, K., 2006. The Rubble-Pile Asteroid Itokawa as Observed by Hayabusa. Science (80-. ). 312, 1330–1334.

Fukuhara, T., Futaguchi, M., Hashimoto, G.L., Horinouchi, T., Imamura, T., Iwagaimi, N., Kouyama, T., Murakami, S.Y., Nakamura, M., Ogohara, K., Sato, M., Sato, T.M., Suzuki, M., Taguchi, M., Takagi, S., Ueno, M., Watanabe, S., Yamada, M., Yamazaki, A., 2017. Large stationary gravity wave in the atmosphere of Venus. Nat. Geosci. 10, 85–88.

Galiano, A., Palomba, E., D'Amore, M., Zinzi, A., Dirri, F., Longobardo, A., Kitazato, K.,

Iwata, T., Matsuoka, M., Hiroi, T., Takir, D., Nakamura, T., Abe, M., Ohtake, M., Matsuura, S., Watanabe, S., Yoshikawa, M., Saiki, T., Tanaka, S., Okada, T., Yamamoto, Y., Takei, Y., Shirai, K., Hirata, N., Matsumoto, K., Tsuda, Y., 2020. Characterization of the Ryugu surface by means of the variability of the near-infrared spectral slope in NIRS3 data. Icarus 351.

Giblin, I., Martelli, G., Farinella, P., Paolicchi, P., Di Martino, M., Smith, P.N., 1998. The Properties of Fragments from Catastrophic Disruption Events. Icarus 134, 77–112.

Goldreich, P., Ward, W.R., 1973. The Formation of Planetismals. Astrophys. J. 183, 1051–1061.

Grott, M., Knollenberg, J., Borgs, B., Hänschke, F., Kessler, E., Helbert, J., Maturilli, A., Müller, N., 2017. The MASCOT Radiometer MARA for the Hayabusa 2 Mission. Space Sci. Rev. 208, 413–431.

Grott, M., Knollenberg, J., Hamm, M., Ogawa, K., Jaumann, R., Otto, K.A., Delbo, M., Michel, P., Biele, J., Neumann, W., Knapmeyer, M., Kührt, E., Senshu, H., Okada, T., Helbert, J., Maturilli, A., Müller, N., Hagermann, A., Sakatani, N., Tanaka, S., Arai, T., Mottola, S., Tachibana, S., Pelivan, I., Drube, L., Vincent, J.B., Yano, H., Pilorget, C., Matz, K.D., Schmitz, N., Koncz, A., Schröder, S.E., Trauthan, F., Schlotterer, M., Krause, C., Ho, T.M., Moussi-Soffys, A., 2019. Low thermal conductivity boulder with high porosity identified on C-type asteroid (162173) Ryugu. Nat. Astron. 3, 971–976.

Harris, A.W., D'Abramo, G., 2015. The population of near-Earth asteroids. Icarus 257, 302–312.

Hartman, W.K., Davis, D.R., 1975. Satellite-Sized Planetesimals and Lunar Origin 515, 504–515.

Hartzell, C.M., 2019. Dynamics of 2D electrostatic dust levitation at asteroids. Icarus 333, 234–242.

Hartzell, C.M., Scheeres, D.J., 2011. The role of cohesive forces in particle launching on the Moon and asteroids. Planet. Space Sci. 59, 1758–1768.

Herčík, D., Auster, H.U., Blum, J., Fornaçon, K.H., Fujimoto, M., Gebauer, K., Güttler, C., Hillenmaier, O., Hördt, A., Liebert, E., Matsuoka, A., Nomura, R., Richter, I., Stoll, B., Weiss, B.P., Glassmeier, K.H., 2017. The MASCOT Magnetometer. Space Sci. Rev. 208, 433–449.

Hercik, D., Auster, H.U., Constantinescu, D., Blum, J., Fornaçon, K.H., Fujimoto, M., Gebauer, K., Grundmann, J.T., Güttler, C., Hillenmaier, O., Ho, T.M., Hördt, A., Krause, C., Kührt, E., Lorda, L., Matsuoka, A., Motschmann, U., Moussi-Soffys, A., Richter, I., Sasaki, K., Scholten, F., Stoll, B., Weiss, B.P., Wolff, F., Glassmeier, K.H., 2020. Magnetic Properties of Asteroid (162173) Ryugu. J. Geophys. Res. Planets 125, 1–11.

Hergenrother, C.W., Whiteley, R.J., 2011. A survey of small fast rotating asteroids among the near-Earth asteroid population. Icarus 214, 194–209.

Hirabayashi, M., Davis, A.B., Fahnestock, E.G., Richardson, D.C., Michel, P., Cheng, A.F., Rivkin, A.S., Scheeres, D.J., Chesley, S.R., Yu, Y., Naidu, S.P., Schwartz, S.R., Benner, L.A.M., Pravec, P., Stickle, A.M., Jutzi, M., 2019a. Assessing possible mutual orbit period change by shape deformation of Didymos after a kinetic impact in the NASA-led Double Asteroid Redirection Test. Adv. Sp. Res. 63, 2515–2534.

Hirabayashi, M., Nakano, R., Tatsumi, E., Walsh, K.J., Barnouin, O.S., Others, 2020. Spin-driven evolution of asteroids' top-shapes at fast and slow spins seen from (101955) Bennu and (162173) Ryugu. Icarus 113946.

Hirabayashi, M., Scheeres, D.J., 2015. Stress and failure analysis of rapidly rotating asteroid (29075) 1950 DA. Astrophys. J. Lett. 798.

Hirabayashi, M., Scheeres, D.J., 2019. Rotationally induced failure of irregularly shaped asteroids. Icarus 317, 354–364.

Hirabayashi, M., Scheeres, D.J., Sánchez, D.-P., Gabriel, T., 2014. Constraints on the physical properties of main belt comet P/2013 R3 from its breakup event. Astrophys. J. Lett. 789, 1–5.

Hirabayashi, M., Schwartz, S.R., Yu, Y., Davis, A.B., Chesley, S.R., Fahnestock, E.G., Michel, P., Richardson, D.C., Naidu, S.P., Scheeres, D.J., Cheng, A.F., Rivkin, A.S., Benner, L.A.M., 2017. Constraints on the perturbed mutual motion in Didymos due to impact-induced deformation of its primary after the DART impact. Mon. Not. R. Astron. Soc. 472, 1641–1648.

Hirabayashi, M., Tatsumi, E., Miyamoto, H., Komatsu, G., Sugita, S., Watanabe, S., Scheeres, D.J., Barnouin, O.S., Michel, P., Honda, C., Michikami, T., Cho, Y., Morota, T., Hirata, Naru, Hirata, Naoyuki, Sakatani, N., Schwartz, S.R., Honda, R., Yokota, Y., Kameda, S., Suzuki, H., Kouyama, T., Hayakawa, M., Matsuoka, M., Yoshioka, K., Ogawa, K., Sawada, H., Yoshikawa, M., Tsuda, Y., 2019b. The Western Bulge of 162173 Ryugu Formed as a Result of a Rotationally Driven Deformation Process. Astrophys. J. 874, L10.

Hirata, N., Morota, T., Cho, Y., Kanamaru, M., Watanabe, S., Sugita, S., Hirata, N., Yamamoto, Y., Noguchi, R., Shimaki, Y., Tatsumi, E., Yoshioka, K., Sawada, H., Yokota, Y., Sakatani, N., Hayakawa, M., Matsuoka, M., Honda, R., Kameda, S., Yamada, M., Kouyama, T., Suzuki, H., Honda, C., Ogawa, K., Tsuda, Y., Yoshikawa, M., Saiki, T., Tanaka, S., Terui, F., Nakazawa, S., Kikuchi, S., Yamaguchi, T., Ogawa, N., Ono, G., Mimasu, Y., Yoshikawa, K., Takahashi, T., Takei, Y., Fujii, A., Takeuchi, H., Okada, T., Shirai, K., Iijima, Y. ichi, 2020. The spatial distribution of impact craters on Ryugu. Icarus 338, 113527.

Ho, T.M., Baturkin, V., Grimm, C., Grundmann, J.T., Hobbie, C., Ksenik, E., Lange, C., Sasaki, K., Schlotterer, M., Talapina, M., Termtanasombat, N., Wejmo, E., Witte, L., Wrasmann, M., Wübbels, G., Rößler, J., Ziach, C., Findlay, R., Biele, J., Krause, C., Ulamec, S., Lange, M., Mierheim, O., Lichtenheldt, R., Maier, M., Reill, J., Sedlmayr, H.J., Bousquet, P., Bellion, A., Bompis, O., Cenac-Morthe, C., Deleuze, M., Fredon, S., Jurado, E., Canalias, E., Jaumann, R., Bibring, J.P., Glassmeier, K.H., Hercik, D., Grott, M., Celotti, L., Cordero, F., Hendrikse, J., Okada, T., 2017. MASCOT—The Mobile Asteroid Surface Scout Onboard the Hayabusa2 Mission. Space Sci. Rev. 208, 339–374.

Holsapple, K.A., 2007. Spin limits of Solar System bodies: From the small fast-rotators to 2003 EL61. Icarus 187, 500–509.

Horinouchi, T., Hayashi, Y., Watanabe, S., Yamada, M., Yamazaki, A., Kouyama, T.,

Taguchi, M., Fukuhara, T., Takagi, M., Ogohara, K., Murakami, S., Peralta, J., Limaye, S.S., Imamura, T., Nakamura, M., Sato, T.M., Satoh, T., 2020. How waves and turbulence maintain the super-rotation of Venus' atmosphere. Science (80-. ). 368, 405–409.

Horinouchi, T., Murakami, S.Y., Satoh, T., Peralta, J., Ogohara, K., Kouyama, T., Imamura, T., Kashimura, H., Limaye, S.S., McGouldrick, K., Nakamura, M., Sato, T.M., Sugiyama, K.I., Takagi, M., Watanabe, S., Yamada, M., Yamazaki, A., Young, E.F., 2017. Equatorial jet in the lower to middle cloud layer of Venus revealed by Akatsuki. Nat. Geosci. 10, 646–651.

Huebner, W.F., Johnson, L.N., Boice, D.C., Bradley, P., Chocron, S., Ghosh, A., Giguere, P.T., Goldstein, R., Guzik, J.A., Keady, J.J., Mukherjee, J., Patrick, W., Plesko, C., Walker, J.D., Wohletz, K., 2009. A comprehensive program for countermeasures against potentially hazardous objects (PHOs)1. Sol. Syst. Res. 43, 334–342.

Ishibashi, K., Shirai, K., Ogawa, K., Wada, K., Honda, R., Arakawa, M., Sakatani, N., Ikeda, Y., 2017. Performance of Hayabusa2 DCAM3-D Camera for Short-Range Imaging of SCI and Ejecta Curtain Generated from the Artificial Impact Crater Formed on Asteroid 162137 Ryugu (1999 JU 3). Space Sci. Rev. 208, 213–238.

Iwata, T., Kitazato, K., Abe, M., Ohtake, M., Arai, Takehiko, Arai, Tomoko, Hirata, N., Hiroi, T., Honda, C., Imae, N., Komatsu, M., Matsunaga, T., Matsuoka, M., Matsuura, S., Nakamura, T., Nakato, A., Nakauchi, Y., Osawa, T., Senshu, H., Takagi, Y., Tsumura, K., Takato, N., Watanabe, S. ichiro, Barucci, M.A., Palomba, E., Ozaki, M., 2017. NIRS3: The Near Infrared Spectrometer on Hayabusa2. Space Sci. Rev. 208, 317–337.

Jaumann, R., Schmitz, N., Ho, T.M., Schröder, S.E., Otto, K.A., Stephan, K., Elgner, S., Krohn, K., Preusker, F., Scholten, F., Biele, J., Ulamec, S., Krause, C., Sugita, S., Matz, K.D., Roatsch, T., Parekh, R., Mottola, S., Grott, M., Michel, P., Trauthan, F., Koncz, A., Michaelis, H., Lange, C., Grundmann, J.T., Maibaum, M., Sasaki, K., Wolff, F., Reill, J., Moussi-Soffys, A., Lorda, L., Neumann, W., Vincent, J.B., Wagner, R., Bibring, J.P., Kameda, S., Yano, H., Watanabe, S., Yoshikawa, M., Tsuda, Y., Okada, T., Yoshimitsu, T., Mimasu, Y., Saiki, T., Yabuta, H., Rauer, H., Honda, R., Morota, T., Yokota, Y., Kouyama, T., 2019. Images from the surface of asteroid Ryugu show rocks similar to carbonaceous chondrite meteorites. Science (80-. ). 365, 817–820.

Jaumann, R., Schmitz, N., Koncz, A., Michaelis, H., Schroeder, S.E., Mottola, S., Trauthan, F., Hoffmann, H., Roatsch, T., Jobs, D., Kachlicki, J., Pforte, B., Terzer, R., Tschentscher, M., Weisse, S., Mueller, U., Perez-Prieto, L., Broll, B., Kruselburger, A., Ho, T.M., Biele, J., Ulamec, S., Krause, C., Grott, M., Bibring, J.P., Watanabe, S., Sugita, S., Okada, T., Yoshikawa, M., Yabuta, H., 2017. The Camera of the MASCOT Asteroid Lander on Board Hayabusa 2. Space Sci. Rev. 208, 375–400.

Johansen, A., Oishi, J.S., Low, M.M. Mac, Klahr, H., Henning, T., Youdin, A., 2007. Rapid planetesimal formation in turbulent circumstellar disks. Nature 448, 1022–1025.

Kadono, T., Arakawa, M., Ito, T., Ohtsuki, K., 2009. Spin rates of fast-rotating asteroids and fragments in impact disruption. Icarus 200, 694–697.

Kameda, S., Suzuki, H., Cho, Y., Koga, S., Yamada, M., Nakamura, T., Hiroi, T., Sawada, H., Honda, R., Morota, T., Honda, C., Takei, A., Takamatsu, T., Okumura, Y., Sato, M., Yasuda, T., Shibasaki, K., Ikezawa, S., Sugita, S., 2015. Detectability of hydrous minerals using ONC-T camera onboard the Hayabusa2 spacecraft. Adv. Sp. Res. 56, 1519–1524.

Kameda, S., Suzuki, H., Takamatsu, T., Cho, Y., Yasuda, T., Yamada, M., Sawada, H., Honda, R., Morota, T., Honda, C., Sato, M., Okumura, Y., Shibasaki, K., Ikezawa, S., Sugita, S., 2017. Preflight Calibration Test Results for Optical Navigation Camera Telescope (ONC-T) Onboard the Hayabusa2 Spacecraft. Space Sci. Rev. 208, 17–31.

Keller, K.U., Barbieri, C., Koschny, D., Lamy, P., Rickman, H., Others, 2010. E-Type Asteroid (2867) Steins as Imaged by OSIRIS on Board Rosetta. Science (80-. ). 327, 190–194.

Kikuchi, S., Terui, F., Ogawa, N., Saiki, T., Ono, G., Yoshikawa, K., Takei, Y., Mimasu, Y., Ikeda, H., Sawada, H., Wal, S. Van, Sugita, S., Watanabe, S., Tsuda, Y., 2020. Design and Reconstruction of the Hayabusa2 Precision Landing on Ryugu. J. Spacecr. Rockets 1–28.

Kitazato, K., Milliken, R.E., Iwata, T., Abe, M., Ohtake, M., Matsuura, S., Arai, T., Nakauchi, Y., Nakamura, T., Matsuoka, M., Senshu, H., Hirata, N., Hiroi, T., Pilorget, C., Brunetto, R., Poulet, F., Riu, L., Bibring, J.P., Takir, D., Domingue, D.L., Vilas, F., Barucci, M.A., Perna, D., Palomba, E., Galiano, A., Tsumura, K., Osawa, T., Komatsu, M., Nakato, A., Arai, T., Takato, N., Matsunaga, T., Takagi, Y., Matsumoto, K., Kouyama, T., Yokota, Y., Tatsumi, E., Sakatani, N., Yamamoto, Y., Okada, T., Sugita, S., Honda, R., Morota, T., Kameda, S., Sawada, H., Honda, C., Yamada, M., Suzuki, H., Yoshioka, K., Hayakawa, M., Ogawa, K., Cho, Y., Shirai, K., Shimaki, Y., Hirata, N., Yamaguchi, A., Ogawa, N., Terui, F., Yamaguchi, T., Takei, Y., Saiki, T., Nakazawa, S., Tanaka, S., Yoshikawa, M., Watanabe, S., Tsuda, Y., 2019. The surface composition of asteroid 162173 Ryugu from Hayabusa2 near-infrared spectroscopy. Science (80-. ). 364, 272–275.

Kouyama, T., Imamura, T., Nakamura, M., Satoh, T., Futaana, Y., 2012. Horizontal structure of planetary-scale waves at the cloud top of Venus deduced from Galileo SSI images with an improved cloud-tracking technique. Planet. Space Sci. 60, 207–216.

Kouyama, T., Imamura, T., Nakamura, M., Satoh, T., Futaana, Y., 2013. Long-term variation in the cloud-tracked zonal velocities at the cloud top of Venus deduced from Venus Express VMC images. J. Geophys. Res. E Planets 118, 37–46.

Kouyama, T., Taguchi, M., Fukuhara, T., Imamura, T., Horinouchi, T., Sato, T.M., Murakami, S., Hashimoto, G.L., Lee, Y.J., Futaguchi, M., Yamada, T., Akiba, M., Satoh, T., Nakamura, M., 2019. Global Structure of Thermal Tides in the Upper Cloud Layer of Venus Revealed by LIR on Board Akatsuki. Geophys. Res. Lett. 46, 9457–9465.

Kouyama, T., Tatsumi, E., Yokota, Y., Yumoto, K., Yamada, M., Honda, R., Kamdeda,

S., Suzuki, H., Sakatani, N., Hayakawa, M., Morota, T., Matsuoka, M., Cho, Y., Honda, C., Sawada, H., Yoshida, K., Sugita, S., 2020. Post-arrival calibration of Hayabusa2's optical navigation cameras (ONCs): Severe effects from touchdown events. Icarus Submitted.

Lauretta, D.S., DellaGiustina, D.N., Bennett, C.A., Golish, D.R., Becker, K.J., Balram-Knutson, S.S., Barnouin, O.S., Becker, T.L., Bottke, W.F., Boynton, W. V., Campins, H., Clark, B.E., Connolly, H.C., Drouet d'Aubigny, C.Y., Dworkin, J.P., Emery, J.P., Enos, H.L., Hamilton, V.E., Hergenrother, C.W., Howell, E.S., Izawa, M.R.M., Kaplan, H.H., Nolan, M.C., Rizk, B., Roper, H.L., Scheeres, D.J., Smith, P.H., Walsh, K.J., Wolner, C.W.V., Highsmith, D.E., Small, J., Vokrouhlický, D., Bowles, N.E., Brown, E., Donaldson Hanna, K.L., Warren, T., Brunet, C., Chicoine, R.A., Desjardins, S., Gaudreau, D., Haltigin, T., Millington-Veloza, S., Rubi, A., Aponte, J., Gorius, N., Lunsford, A., Allen, B., Grindlay, J., Guevel, D., Hoak, D., Hong, J., Schrader, D.L., Bayron, J., Golubov, O., Sánchez, P., Stromberg, J., Hirabayashi, M., Hartzell, C.M., Oliver, S., Rascon, M., Harch, A., Joseph, J., Squyres, S., Richardson, D., McGraw, L., Ghent, R., Binzel, R.P., Asad, M.M.A., Johnson, C.L., Philpott, L., Susorney, H.C.M., Cloutis, E.A., Hanna, R.D., Ciceri, F., Hildebrand, A.R., Ibrahim, E.M., Breitenfeld, L., Glotch, T., Rogers, A.D., Ferrone, S., Thomas, C.A., Fernandez, Y., Chang, W., Cheuvront, A., Trang, D., Tachibana, S., Yurimoto, H., Brucato, J.R., Poggiali, G., Pajola, M., Dotto, E., Epifani, E.M., Crombie, M.K., Lantz, C., de Leon, J., Licandro, J., Garcia, J.L.R., Clemett, S., Thomas-Keprta, K., Van wal, S., Yoshikawa, M., Bellerose, J., Bhaskaran, S., Boyles, C., Chesley, S.R., Elder, C.M., Farnocchia, D., Harbison, A., Kennedy, B., Knight, A., Martinez-Vlasoff, N., Mastrodemos, N., McElrath, T., Owen, W., Park, R., Rush, B., Swanson, L., Takahashi, Y., Velez, D., Yetter, K., Thayer, C., Adam, C., Antreasian, P., Bauman, J., Bryan, C., Carcich, B., Corvin, M., Geeraert, J., Hoffman, J., Leonard, J.M., Lessac-Chenen, E., Levine, A., McAdams, J., McCarthy, L., Nelson, D., Page, B., Pelgrift, J., Sahr, E., Stakkestad, K., Stanbridge, D., Wibben, D., Williams, B., Williams, K., Wolff, P., Hayne, P., Kubitschek, D., Barucci, M.A., Deshapriya, J.D.P., Fornasier, S., Fulchignoni, M., Hasselmann, P., Merlin, F., Praet, A., Bierhaus, E.B., Billett, O., Boggs, A., Buck, B., Carlson-Kelly, S., Cerna, J., Chaffin, K., Church, E., Coltrin, M., Daly, A., Deguzman, A., Dubisher, R., Eckart, D., Ellis, D., Falkenstern, P., Fisher, A., Fisher, M.E., Fleming, P., Fortney, K., Francis, S., Freund, S., Gonzales, S., Haas, P., Hasten, A., Hauf, D., Hilbert, A., Howell, D., Jaen, F., Jayakody, N., Jenkins, M., Johnson, K., Lefevre, M., Ma, H., Mario, C., Martin, K., May, C., McGee, M., Miller, B., Miller, C., Miller, G., Mirfakhrai, A., Muhle, E., Norman, C., Olds, R., Parish, C., Ryle, M., Schmitzer, M., Sherman, P., Skeen, M., Susak, M., Sutter, B., Tran, Q., Welch, C., Witherspoon, R., Wood, J., Zareski, J., Arvizu-Jakubicki, M., Asphaug, E., Audi, E., Ballouz, R.L., Bandrowski, R., Bendall, S., Bloomenthal, H., Blum, D., Brodbeck, J., Burke, K.N., Chojnacki, M., Colpo, A., Contreras, J., Cutts, J., Dean, D., Diallo, B., Drinnon, D., Drozd, K., Enos, R., Fellows, C., Ferro, T., Fisher, M.R., Fitzgibbon, G., Fitzgibbon, M., Forelli, J., Forrester, T., Galinsky, I., Garcia, R., Gardner, A., Habib, N., Hamara, D., Hammond, D., Hanley, K., Harshman, K.,

Herzog, K., Hill, D., Hoekenga, C., Hooven, S., Huettner, E., Janakus, A., Jones, J., Kareta, T.R., Kidd, J., Kingsbury, K., Koelbel, L., Kreiner, J., Lambert, D., Lewin, C., Lovelace, B., Loveridge, M., Lujan, M., Maleszewski, C.K., Malhotra, R., Marchese, K., McDonough, E., Mogk, N., Morrison, V., Morton, E., Munoz, R., Nelson, J., Padilla, J., Pennington, R., Polit, A., Ramos, N., Reddy, V., Riehl, M., Salazar, S., Schwartz, S.R., Selznick, S., Shultz, N., Stewart, S., Sutton, S., Swindle, T., Tang, Y.H., Westermann, M., Worden, D., Zega, T., Zeszut, Z., Bjurstrom, A., Bloomquist, L., Dickinson, C., Keates, E., Liang, J., Nifo, V., Taylor, A., Teti, F., Caplinger, M., Bowles, H., Carter, S., Dickenshied, S., Doerres, D., Fisher, T., Hagee, W., Hill, J., Miner, M., Noss, D., Piacentine, N., Smith, M., Toland, A., Wren, P., Bernacki, M., Munoz, D.P., Watanabe, S.I., Sandford, S.A., Aqueche, A., Ashman, B., Barker, M., Bartels, A., Berry, K., Bos, B., Burns, R., Calloway, A., Carpenter, P., Castro, N., Cosentino, R., Donaldson, J., Cook, J.E., Emr, C., Everett, D., Fennell, D., Fleshman, K., Folta, D., Gallagher, D., Garvin, J., Getzandanner, K., Glavin, D., Hull, S., Hyde, K., Ido, H., Ingegneri, A., Jones, N., Kaotira, P., Lim, L.F., Liounis, A., Lorentson, C., Lorenz, D., Lyzhoft, J., Mazarico, E.M., Mink, R., Moore, W., Moreau, M., Mullen, S., Nagy, J., Neumann, G., Nuth, J., Poland, D., Reuter, D.C., Rhoads, L., Rieger, S., Rowlands, D., Sallitt, D., Scroggins, A., Shaw, G., Simon, A.A., Swenson, J., Vasudeva, P., Wasser, M., Zellar, R., Grossman, J., Johnston, G., Morris, M., Wendel, J., Burton, A., Keller, L.P., McNamara, L., Messenger, S., Nakamura-Messenger, K., Nguyen, A., Righter, K., Queen, E., Bellamy, K., Dill, K., Gardner, S., Giuntini, M., Key, B., Kissell, J., Patterson, D., Vaughan, D., Wright, B., Gaskell, R.W., Le Corre, L., Li, J.Y., Molaro, J.L., Palmer, E.E., Siegler, M.A., Tricarico, P., Weirich, J.R., Zou, X.D., Ireland, T., Tait, K., Bland, P., Anwar, S., Bojorquez-Murphy, N., Christensen, P.R., Haberle, C.W., Mehall, G., Rios, K., Franchi, I., Rozitis, B., Beddingfield, C.B., Marshall, J., Brack, D.N., French, A.S., McMahon, J.W., Jawin, E.R., McCoy, T.J., Russell, S., Killgore, M., Bandfield, J.L., Clark, B.C., Chodas, M., Lambert, M., Masterson, R.A., Daly, M.G., Freemantle, J., Seabrook, J.A., Craft, K., Daly, R.T., Ernst, C., Espiritu, R.C., Holdridge, M., Jones, M., Nair, A.H., Nguyen, L., Peachey, J., Perry, M.E., Plescia, J., Roberts, J.H., Steele, R., Turner, R., Backer, J., Edmundson, K., Mapel, J., Milazzo, M., Sides, S., Manzoni, C., May, B., Delbo', M., Libourel, G., Michel, P., Ryan, A., Thuillet, F., Marty, B., 2019a. The unexpected surface of asteroid (101955) Bennu. Nature 568, 55–60.

Lauretta, D.S., Hergenrother, C.W., Chesley, S.R., Leonard, J.M., Pelgrift, J.Y., Adam, C.D., Asad, M. Al, Antreasian, P.G., Ballouz, R.L., Becker, K.J., Bennett, C.A., Bos, B.J., Bottke, W.F., Brozović, M., Campins, H., Connolly, H.C., Daly, M.G., Davis, A.B., de León, J., DellaGiustina, D.N., Drouet d'Aubigny, C.Y., Dworkin, J.P., Emery, J.P., Farnocchia, D., Glavin, D.P., Golish, D.R., Hartzell, C.M., Jacobson, R.A., Jawin, E.R., Jenniskens, P., Kidd, J.N., Lessac-Chenen, E.J., Li, J.Y., Libourel, G., Licandro, J., Liounis, A.J., Maleszewski, C.K., Manzoni, C., May, B., McCarthy, L.K., McMahon, J.W., Michel, P., Molaro, J.L., Moreau, M.C., Nelson, D.S., Owen, W.M., Rizk, B., Roper, H.L., Rozitis, B., Sahr, E.M., Scheeres, D.J., Seabrook, J.A., Selznick, S.H., Takahashi, Y., Thuillet, F., Tricarico, P.,

Vokrouhlický, D., Wolner, C.W.V., 2019b. Episodes of particle ejection from the surface of the active asteroid (101955) Bennu. Science (80-. ). 366.

Lee, Y.J., Jessup, K.-L., Perez-Hoyos, S., Titov, D. V., Lebonnois, S., Peralta, J., Horinouchi, T., Imamura, T., Limaye, S., Marcq, E., Takagi, M., Yamazaki, A., Yamada, M., Watanabe, S., Murakami, S., Ogohara, K., McClintock, W.M., Holsclaw, G., Roman, A., 2019. Long-term Variations of Venus's 365 nm Albedo Observed by Venus Express , Akatsuki , MESSENGER , and the Hubble Space Telescope. Astron. J. 158, 126.

Leung, R.Y., Barbee, B.W., Seery, B.D., Bambacus, M., Finewood, L., Greenaugh, K.C., Lewis, A., Dearborn, D.S.P., Miller, P.L., Weaver, R.P., Plesko, C., 2017. Multi-organization - Multi-discipline effort developing a mitigation concept for planetary defense. IEEE Aerosp. Conf. Proc.

Lu, E.T., Love, S.G., 2005. Gravitational tractor for towing asteroids. Nature 438, 177–178.

Mainzer, A., 2017. The future of planetary defense. J. Geophys. Res. Planets 122, 789–793.

McMahon, J.W., Scheeres, D.J., 2017. The effect of asteroid topography on surface ablation deflection. Adv. Sp. Res. 59, 1144–1155.

Michel, P., Ballouz, R.L., Barnouin, O.S., Jutzi, M., Walsh, K.J., May, B.H., Manzoni, C., Richardson, D.C., Schwartz, S.R., Sugita, S., Watanabe, S., Miyamoto, H., Hirabayashi, M., Bottke, W.F., Connolly, H.C., Yoshikawa, M., Lauretta, D.S., 2020a. Collisional formation of top-shaped asteroids and implications for the origins of Ryugu and Bennu. Nat. Commun. 11.

Michel, P., Benz, W., Tanga, P., Richardson, D.C., 2001. Collisions and gravitational reaccumulation: Forming asteroid families and satellites. Science (80-. ). 294, 1696–1700.

Michel, P., Cheng, A., Küppers, M., Pravec, P., Blum, J., Delbo, M., Green, S.F., Rosenblatt, P., Tsiganis, K., Vincent, J.B., Biele, J., Ciarletti, V., Hérique, A., Ulamec, S., Carnelli, I., Galvez, A., Benner, L., Naidu, S.P., Barnouin, O.S., Richardson, D.C., Rivkin, A., Scheirich, P., Moskovitz, N., Thirouin, A., Schwartz, S.R., Campo Bagatin, A., Yu, Y., 2016. Science case for the Asteroid Impact Mission (AIM): A component of the Asteroid Impact & Deflection Assessment (AIDA) mission. Adv. Sp. Res. 57, 2529–2547.

Michel, P., Kueppers, M., Camto Bagatin, A., Carry, B., Charnoz, S., de Leon, J., Others, 2020b. Science and planetary defense aspects of the ESA Hera mission. In: 51st Lunar and Planetary Science Conference. p. 1441.

Michel, P., Kueppers, M., Sierks, H., Carnelli, I., Cheng, A.F., Mellab, K., Granvik, M., Kestilä, A., Kohout, T., Muinonen, K., Näsilä, A., Penttila, A., Tikka, T., Tortora, P., Ciarletti, V., Hérique, A., Murdoch, N., Asphaug, E., Rivkin, A., Barnouin, O., Bagatin, A.C., Pravec, P., Richardson, D.C., Schwartz, S.R., Tsiganis, K., Ulamec, S., Karatekin, O., 2018. European component of the AIDA mission to a binary asteroid: Characterization and interpretation of the impact of the DART mission. Adv. Sp. Res. 62, 2261–2272.

Michikami, T., Honda, C., Miyamoto, H., Hirabayashi, M., Hagermann, A., Irie, T.,

Nomura, K., Ernst, C.M., Kawamura, M., Sugimoto, K., Tatsumi, E., Morota, T., Hirata, Naru, Noguchi, T., Cho, Y., Kameda, S., Kouyama, T., Yokota, Y., Noguchi, R., Hayakawa, M., Hirata, Naoyuki, Honda, R., Matsuoka, M., Sakatani, N., Suzuki, H., Yamada, M., Yoshioka, K., Sawada, H., Hemmi, R., Kikuchi, H., Ogawa, K., Watanabe, S. ichiro, Tanaka, S., Yoshikawa, M., Tsuda, Y., Sugita, S., 2019. Boulder size and shape distributions on asteroid Ryugu. Icarus 331, 179–191.

Michikami, T., Nakamura, A.M., Hirata, N., 2010. The shape distribution of boulders on Asteroid 25143 Itokawa: Comparison with fragments from impact experiments. Icarus 207, 277–284.

Mizuno, T., Kase, T., Shiina, T., Mita, M., Namiki, N., Senshu, H., Yamada, R., Noda, H., Kunimori, H., Hirata, N., Terui, F., Mimasu, Y., 2017. Development of the Laser Altimeter (LIDAR) for Hayabusa2. Space Sci. Rev. 208, 33–47.

Molaro, J.L., Hergenrother, C.W., Chesley, S.R., Walsh, K.J., Hanna, R.D., Haberle, C.W., Schwartz, S.R., Ballouz, R. -L., Bottke, W.F., Campins, H.J., Lauretta, D.S., 2020a. Thermal fatigue as a driving mechanism for activity on asteroid Bennu. J. Geophys. Res. Planets 1–24.

Molaro, J.L., Walsh, K.J., Jawin, E.R., Ballouz, R.L., Bennett, C.A., DellaGiustina, D.N., Golish, D.R., Drouet d'Aubigny, C., Rizk, B., Schwartz, S.R., Hanna, R.D., Martel, S.J., Pajola, M., Campins, H., Ryan, A.J., Bottke, W.F., Lauretta, D.S., 2020b. In situ evidence of thermally induced rock breakdown widespread on Bennu's surface. Nat. Commun. 11, 1–11.

Morota, T., Sugita, S., Cho, Y., Kanamaru, M., Tatsumi, E., Sakatani, N., Honda, R., Hirata, N., Kikuchi, H., Yamada, M., Yokota, Y., Kameda, S., Matsuoka, M., Sawada, H., Honda, C., Kouyama, T., Ogawa, K., Suzuki, H., Yoshioka, K., Hayakawa, M., Hirata, N., Hirabayashi, M., Miyamoto, H., Michikami, T., Hiroi, T., Hemmi, R., Barnouin, O.S., Ernst, C.M., Kitazato, K., Nakamura, T., Riu, L., Senshu, H., Kobayashi, H., Sasaki, S., Komatsu, G., Tanabe, N., Fujii, Y., Irie, T., Suemitsu, M., Takaki, N., Sugimoto, C., Yumoto, K., Ishida, M., Kato, H., Moroi, K., Domingue, D., Michel, P., Pilorget, C., Iwata, T., Abe, M., Ohtake, M., Nakauchi, Y., Tsumura, K., Yabuta, H., Ishihara, Y., Noguchi, R., Matsumoto, K., Miura, A., Namiki, N., Tachibana, S., Arakawa, M., Ikeda, H., Wada, K., Mizuno, T., Hirose, C., Hosoda, S., Mori, O., Shimada, T., Soldini, S., Tsukizaki, R., Yano, H., Ozaki, M., Takeuchi, H., Yamamoto, Y., Okada, T., Shimaki, Y., Shirai, K., Iijima, Y., Noda, H., Kikuchi, S., Yamaguchi, T., Ogawa, N., Ono, G., Mimasu, Y., Yoshikawa, K., Takahashi, T., Takei, Y., Fujii, A., Nakazawa, S., Terui, F., Tanaka, S., Yoshikawa, M., Saiki, T., Watanabe, S., Tsuda, Y., 2020. Sample collection from asteroid (162173) Ryugu by Hayabusa2: Implications for surface evolution. Science (80-. ). 368, 654–659.

Nagao, K., Okazaki, R., Nakamura, T., Miura, Y.N., Osawa, T., Bajo, K., Others, 2011. Irradiation History of Itokawa Regolith 333, 1128–1131.

Naidu, S.P., Benner, L.A.M., Brozovic, M., Nolan, M.C., Ostro, S.J., Margot, J.L., Giorgini, J.D., Hirabayashi, T., Scheeres, D.J., Pravec, P., Scheirich, P., Magri, C., Jao, J.S., 2020. Radar observations and a physical model of binary near-Earth asteroid 65803 Didymos, target of the DART mission. Icarus 348, 113777.

Nakano, R., Hirabayashi, M., 2020. Mass-shedding Activities of Asteroid (3200) Phaethon Enhanced by Its Rotation. Astrophys. J. 892, L22.

Nesvorný, D., Vokrouhlický, D., 2007. Analytic Theory of the YORP Effect for Near-Spherical Objects. Astron. J. 134, 1,750-1,768.

Nesvorný, D., Vokrouhlický, D., 2008. Analytic Theory for the Yarkovsky-O'Keefe-Radzievski-Paddack effect on Obliquity. Astron. J. 136, 291–299.

Ogawa, K., Shirai, K., Sawada, H., Arakawa, M., Honda, R., Wada, K., Ishibashi, K., Iijima, Y. ichi, Sakatani, N., Nakazawa, S., Hayakawa, H., 2017. System Configuration and Operation Plan of Hayabusa2 DCAM3-D Camera System for Scientific Observation During SCI Impact Experiment. Space Sci. Rev. 208, 125–142.

Ogawa, N., Terui, F., Mimasu, Y., Yoshikawa, K., Ono, G., Yasuda, S., Matsushima, K., Masuda, T., Hihara, H., Sano, J., Matsuhisa, T., Danno, S., Yamada, M., Yokota, Y., Takei, Y., Saiki, T., Tsuda, Y., 2020. Image-based autonomous navigation of Hayabusa2 using artificial landmarks: The design and brief in-flight results of the first landing on asteroid Ryugu. Astrodynamics 4, 89–103.

Okada, T., Fukuhara, T., Tanaka, S., Taguchi, M., Arai, T., Senshu, H., Sakatani, N., Shimaki, Y., Demura, H., Ogawa, Y., Suko, K., Sekiguchi, T., Kouyama, T., Takita, J., Matsunaga, T., Imamura, T., Wada, T., Hasegawa, S., Helbert, J., Müller, T.G., Hagermann, A., Biele, J., Grott, M., Hamm, M., Delbo, M., Hirata, Naru, Hirata, Naoyuki, Yamamoto, Y., Sugita, S., Namiki, N., Kitazato, K., Arakawa, M., Tachibana, S., Ikeda, H., Ishiguro, M., Wada, K., Honda, C., Honda, R., Ishihara, Y., Matsumoto, K., Matsuoka, M., Michikami, T., Miura, A., Morota, T., Noda, H., Noguchi, R., Ogawa, K., Shirai, K., Tatsumi, E., Yabuta, H., Yokota, Y., Yamada, M., Abe, M., Hayakawa, M., Iwata, T., Ozaki, M., Yano, H., Hosoda, S., Mori, O., Sawada, H., Shimada, T., Takeuchi, H., Tsukizaki, R., Fujii, A., Hirose, C., Kikuchi, S., Mimasu, Y., Ogawa, N., Ono, G., Takahashi, T., Takei, Y., Yamaguchi, T., Yoshikawa, K., Terui, F., Saiki, T., Nakazawa, S., Yoshikawa, M., Watanabe, S., Tsuda, Y., 2020. Highly porous nature of a primitive asteroid revealed by thermal imaging. Nature 579, 518–522.

Okada, T., Fukuhara, T., Tanaka, S., Taguchi, M., Imamura, T., Arai, T., Senshu, H., Ogawa, Y., Demura, H., Kitazato, K., Nakamura, R., Kouyama, T., Sekiguchi, T., Hasegawa, S., Matsunaga, T., Wada, T., Takita, J., Sakatani, N., Horikawa, Y., Endo, K., Helbert, J., Müller, T.G., Hagermann, A., 2017. Thermal Infrared Imaging Experiments of C-Type Asteroid 162173 Ryugu on Hayabusa2. Space Sci. Rev. 208, 255–286.

Okazaki, R., Sawada, H., Yamanouchi, S., Tachibana, S., Miura, Y.N., Sakamoto, K., Takano, Y., Abe, M., Itoh, S., Yamada, K., Yabuta, H., Okamoto, C., Yano, H., Noguchi, T., Nakamura, T., Nagao, K., 2017. Hayabusa2 Sample Catcher and Container: Metal-Seal System for Vacuum Encapsulation of Returned Samples with Volatiles and Organic Compounds Recovered from C-Type Asteroid Ryugu. Space Sci. Rev. 208, 107–124.

Ono, G., Terui, F., Ogawa, N., Mimasu, Y., Yoshikawa, K., Takei, Y., Saiki, T., Tsuda, Y., 2020. Design and flight results of GNC systems in Hayabusa2 descent operations. Astrodynamics 4, 105–117.

Ostro, S.J., Pravec, P., Benner, L.A.M., Hudson, R.S., Šarounová, L., Hicks, M.D., Rabinowitz, D.L., Scotti, J. V., Tholen, D.J., Wolf, M., Jurgens, R.F., Thomas, M.L., Giorgini, J.D., Chodas, P.W., Yeomans, D.K., Rose, R., Frye, R., Rosema, K.D., Winkler, R., Slade, M.A., 1999. Radar and optical observations of asteroid 1998 KY26. Science (80-. ). 285, 557–559.

Pepper, J., Rodriguez, J.E., Collins, K.A., Johnson, J.A., Fulton, B.J., Howard, A.W., Beatty, T.G., Stassun, K.G., Isaacson, H., Colón, K.D., Lund, M.B., Kuhn, R.B., Siverd, R.J., Gaudi, B.S., Tan, T.G., Curtis, I., Stockdale, C., Mawet, D., Bottom, M., James, D., Zhou, G., Bayliss, D., Cargile, P., Bieryla, A., Penev, K., Latham, D.W., Labadie-Bartz, J., Kielkopf, J., Eastman, J.D., Oberst, T.E., Jensen, E.L.N., Nelson, P., Sliski, D.H., Wittenmyer, R.A., McCrady, N., Wright, J.T., Relles, H.M., Stevens, D.J., Joner, M.D., Hintz, E., 2017. KELT-11b: A Highly Inflated Sub-Saturn Exoplanet Transiting the V = 8 Subgiant HD 93396 . Astron. J. 153, 215.

Pilorget, C., Fernando, J., Riu, L., Kitazato, K., Iwata, T., 2020. Global-scale albedo and spectro-photometric properties of Ryugu from NIRS3/Hayabusa2, implications for the composition of Ryugu and the representativity of the returned samples. Icarus In Press, 109231.

Poppe, A.R., Lisse, C.M., Piquette, M., Zemcov, M., Horányi, M., James, D., Szalay, J.R., Bernardoni, E., Stern, S.A., 2019. Constraining the Solar System's Debris Disk with In Situ New Horizons Measurements from the Edgeworth–Kuiper Belt . Astrophys. J. 881, L12.

Raducan, S.D., Davison, T.M., Collins, G.S., 2020. The effects of asteroid layering on ejecta mass-velocity distribution and implications for impact momentum transfer. Planet. Space Sci. 180, 104756.

Raducan, S.D., Davison, T.M., Luther, R., Collins, G.S., 2019. The role of asteroid strength, porosity and internal friction in impact momentum transfer. Icarus 329, 282–295.

Rozitis, B., Maclennan, E., Emery, J.P., 2014. Cohesive forces prevent the rotational breakup of rubble-pile asteroid (29075) 1950 da. Nature 512, 174–176.

Saiki, T., Takei, Y., Mimasu, Y., Sawada, H., Ogawa, N., Ono, G., Yoshikawa, K., Terui, F., Arakawa, M., Sugita, S., Watanabe, S. ichiro, Yoshikawa, M., Nakazawa, S., Tsuda, Y., 2020. Hayabusa2's kinetic impact experiment: Operational planning and results. Acta Astronaut. 175, 362–374.

Sarli, B.V., Tsuda, Y., 2017. Hayabusa 2 extension plan: Asteroid selection and trajectory design. Acta Astronaut. 138, 225–232.

Sawada, H., Ogawa, K., Shirai, K., Kimura, S., Hiromori, Y., Mimasu, Y., 2017a. Deployable Camera (DCAM3) System for Observation of Hayabusa2 Impact Experiment. Space Sci. Rev. 208, 143–164.

Sawada, H., Okazaki, R., Tachibana, S., Sakamoto, K., Takano, Y., Okamoto, C., Yano, H., Miura, Y., Abe, M., Hasegawa, S., Noguchi, T., 2017b. Hayabusa2 Sampler: Collection of Asteroidal Surface Material. Space Sci. Rev. 208, 81–106.

Scheeres, D.J., Hartzell, C.M., Sánchez, P., Swift, M., 2010. Scaling forces to asteroid surfaces: The role of cohesion. Icarus 210, 968–984.

Scholten, F., Preusker, F., Elgner, S., Matz, K.-D., Jaumann, R., Hamm, M., Schröder, S.E., Koncz, A., Schmitz, N., Trauthan, F., Grott, M., Biele, J., Ho, T.-M., Kameda, S., Sugita, S., 2019. The Hayabusa2 lander MASCOT on the surface of asteroid (162173) Ryugu – Stereo-photogrammetric analysis of MASCam image data. Astron. Astrophys. 632, L5.

Senshu, H., Oshigami, S., Kobayashi, M., Yamada, R., Namiki, N., Noda, H., Ishihara, Y., Mizuno, T., 2017. Dust Detection Mode of the Hayabusa2 LIDAR. Space Sci. Rev. 208, 65–79.

Shimaki, Y., Senshu, H., Sakatani, N., Okada, T., Fukuhara, T., Tanaka, S., Taguchi, M., Arai, T., Demura, H., Ogawa, Y., Suko, K., Sekiguchi, T., Kouyama, T., Hasegawa, S., Takita, J., Matsunaga, T., Imamura, T., Wada, T., Kitazato, K., Hirata, Naru, Hirata, Naoyuki, Noguchi, R., Sugita, S., Kikuchi, S., Yamaguchi, T., Ogawa, N., Ono, G., Mimasu, Y., Yoshikawa, K., Takahashi, T., Takei, Y., Fujii, A., Takeuchi, H., Yamamoto, Y., Yamada, M., Shirai, K., Iijima, Y. ichi, Ogawa, K., Nakazawa, S., Terui, F., Saiki, T., Yoshikawa, M., Tsuda, Y., Watanabe, S. ichiro, 2020. Thermophysical properties of the surface of asteroid 162173 Ryugu: Infrared observations and thermal inertia mapping. Icarus 348, 113835.

Siverd, R.J., Collins, K.A., Zhou, G., Quinn, S.N., Gaudi, B.S., Stassun, K.G., Johnson, M.C., Bieryla, A., Latham, D.W., Ciardi, D.R., Rodriguez, J.E., Penev, K., Pinsonneault, M., Pepper, J., Eastman, J.D., Relles, H., Kielkopf, J.F., Gregorio, J., Oberst, T.E., Aldi, G.F., Esquerdo, G.A., Calkins, M.L., Berlind, P., Dressing, C.D., Patel, R., Stevens, D.J., Beatty, T.G., Lund, M.B., Labadie-Bartz, J., Kuhn, R.B., Colón, K.D., James, D., Yao, X., Johnson, J.A., Wright, J.T., McCrady, N., Wittenmyer, R.A., Johnson, S.A., Sliski, D.H., Jensen, E.L.N., Cohen, D.H., McLeod, K.K., Penny, M.T., Joner, M.D., Stephens, D.C., Villanueva, S., Zambelli, R., Stockdale, C., Evans, P., Tan, T.-G., Curtis, I.A., Reed, P.A., Trueblood, M., Trueblood, P., 2017. KELT-19Ab: A P ~ 4.6-day Hot Jupiter Transiting a Likely Am Star with a Distant Stellar Companion. Astron. J. 155, 35.

Stickle, A.M., Bruck Syal, M., Cheng, A.F., Collins, G.S., Davison, T.M., Gisler, G., Güldemeister, N., Heberling, T., Luther, R., Michel, P., Miller, P., Owen, J.M., Rainey, E.S.G., Rivkin, A.S., Rosch, T., Wünnemann, K., 2020. Benchmarking impact hydrocodes in the strength regime: Implications for modeling deflection by a kinetic impactor. Icarus 338, 113446.

Sugita, S., Honda, R., Morota, T., Kameda, S., Sawada, H., Tatsumi, E., Yamada, M., Honda, C., Yokota, Y., Kouyama, T., Sakatani, N., Ogawa, K., Suzuki, H., Okada, T., Namiki, N., Tanaka, S., Iijima, Y., Yoshioka, K., Hayakawa, M., Cho, Y., Matsuoka, M., Hirata, N., Hirata, N., Miyamoto, H., Domingue, D., Hirabayashi, M., Nakamura, T., Hiroi, T., Michikami, T., Michel, P., Ballouz, R.L., Barnouin, O.S., Ernst, C.M., Schröder, S.E., Kikuchi, H., Hemmi, R., Komatsu, G., Fukuhara, T., Taguchi, M., Arai, T., Senshu, H., Demura, H., Ogawa, Y., Shimaki, Y., Sekiguchi, T., Müller, T.G., Hagermann, A., Mizuno, T., Noda, H., Matsumoto, K., Yamada, R., Ishihara, Y., Ikeda, H., Araki, H., Yamamoto, K., Abe, S., Yoshida, F., Higuchi, A., Sasaki, S., Oshigami, S., Tsuruta, S., Asari, K., Tazawa, S., Shizugami, M., Kimura, J., Otsubo, T., Yabuta, H., Hasegawa, S., Ishiguro, M., Tachibana, S.,

Palmer, E., Gaskell, R., Le Corre, L., Jaumann, R., Otto, K., Schmitz, N., Abell, P.A., Barucci, M.A., Zolensky, M.E., Vilas, F., Thuillet, F., Sugimoto, C., Takaki, N., Suzuki, Y., Kamiyoshihara, H., Okada, M., Nagata, K., Fujimoto, M., Yoshikawa, M., Yamamoto, Y., Shirai, K., Noguchi, R., Ogawa, N., Terui, F., Kikuchi, S., Yamaguchi, T., Oki, Y., Takao, Y., Takeuchi, H., Ono, G., Mimasu, Y., Yoshikawa, K., Takahashi, T., Takei, Y., Fujii, A., Hirose, C., Nakazawa, S., Hosoda, S., Mori, O., Shimada, T., Soldini, S., Iwata, T., Abe, M., Yano, H., Tsukizaki, R., Ozaki, M., Nishiyama, K., Saiki, T., Watanabe, S., Tsuda, Y., 2019. The geomorphology, color, and thermal properties of Ryugu: Implications for parent-body processes. Science (80-. ). 364.

Suzuki, H., Yamada, M., Kouyama, T., Tatsumi, E., Kameda, S., Honda, R., Sawada, H., Ogawa, N., Morota, T., Honda, C., Sakatani, N., Hayakawa, M., Yokota, Y., Yamamoto, Y., Sugita, S., 2018. Initial inflight calibration for Hayabusa2 optical navigation camera (ONC) for science observations of asteroid Ryugu. Icarus 300, 341–359.

Takita, J., Senshu, H., Tanaka, S., 2017. Feasibility and Accuracy of Thermophysical Estimation of Asteroid 162173 Ryugu (1999 JU3) from the Hayabusa2 Thermal Infrared Imager. Space Sci. Rev. 208, 287–315.

Tatsumi, E., Domingue, D., Yokota, Y., Schroder, S., Hasegawa, S., Kuroda, D., Ishiguro, M., Hiroi, T., Honda, R., Hemmi, R., Le Corre, L., Sakatani, N., Morota, T., Yamada, M., Kameda, S., Koyama, T., Suzuki, H., Cho, Y., Yoshioka, K., Matsuoka, M., Honda, C., 2020a. Global photometric properties of (162173) Ryugu. Astron. Astrophys. 83.

Tatsumi, E., Kouyama, T., Suzuki, H., Yamada, M., Sakatani, N., Kameda, S., Yokota, Y., Honda, R., Morota, T., Moroi, K., Tanabe, N., Kamiyoshihara, H., Ishida, M., Yoshioka, K., Sato, H., Honda, C., Hayakawa, M., Kitazato, K., Sawada, H., Sugita, S., 2019. Updated inflight calibration of Hayabusa2's optical navigation camera (ONC) for scientific observations during the cruise phase. Icarus 325, 153–195.

Tatsumi, E., Sugimoto, C., Riu, L., Sugita, S., Nakamura, T., Hiroi, T., Morota, T., Popescu, M., Michikami, T., Kitazato, K., Matsuoka, M., Kameda, S., Honda, R., Others, 2020b. Collisional history of Ryugu's parent body from bright surface boulders. Nat. Astron.

Tsuda, Y., Takeuchi, H., Ogawa, N., Ono, G., Kikuchi, S., Oki, Y., Ishiguro, M., Kuroda, D., Urakawa, S., Okumura, S. ichiro, 2020. Rendezvous to asteroid with highly uncertain ephemeris: Hayabusa2's Ryugu-approach operation result. Astrodynamics 4, 137–147.

Tsuda, Y., Yoshikawa, M., Saiki, T., Nakazawa, S., Watanabe, S. ichiro, 2019. Hayabusa2–Sample return and kinetic impact mission to near-earth asteroid Ryugu. Acta Astronaut. 156, 387–393.

Usui, T., Bajo, K. ichi, Fujiya, W., Furukawa, Y., Koike, M., Miura, Y.N., Sugahara, H., Tachibana, S., Takano, Y., Kuramoto, K., 2020. The Importance of Phobos Sample Return for Understanding the Mars-Moon System. Space Sci. Rev. 216.

Venditti, F.C.F., Marchi, L.O., Misra, A.K., Sanchez, D.M., Prado, A.F.B.A., 2020. Dynamics of tethered asteroid systems to support planetary defense. Eur. Phys. J.

Spec. Top. 229, 1463–1477.

Veverka, J., Thomas, P., Harch, A., Clark, B., Bell, J.F.I., Malin, M., Mcfadden, L.A., Murchie, S., Hawkins, S.E.I., 1997. NEAR's Flyby of 253 Mathilde: Images ofaCAsteroid. Science (80-. ). 278, 2109–2114.

Walsh, K.J., 2018. Rubble Pile Asteroids. Annu. Rev. Astron. Astrophys. 56, 593–624.

Walsh, K.J., Jawin, E.R., Ballouz, R.L., Barnouin, O.S., Bierhaus, E.B., Connolly, H.C., Molaro, J.L., McCoy, T.J., Delbo', M., Hartzell, C.M., Pajola, M., Schwartz, S.R., Trang, D., Asphaug, E., Becker, K.J., Beddingfield, C.B., Bennett, C.A., Bottke, W.F., Burke, K.N., Clark, B.C., Daly, M.G., Dellagiustina, D.N., Dworkin, J.P., Elder, C.M., Golish, D.R., Hildebrand, A.R., Malhotra, R., Marshall, J., Michel, P., Nolan, M.C., Perry, M.E., Rizk, B., Ryan, A., Sandford, S.A., Scheeres, D.J., Susorney, H.C.M., Thuillet, F., Lauretta, D.S., 2019. Craters, boulders and regolith of (101955) bennu indicative of an old and dynamic surface. Nat. Geosci. 12, 242–246.

Warner, B.D., Harris, A.W., Pravec, P., 2009. The asteroid lightcurve database. Icarus 202, 134–146.

Watanabe, S., Hirabayashi, M., Hirata, N., Hirata, N., Noguchi, R., Shimaki, Y., Ikeda, H., Tatsumi, E., Yoshikawa, M., Kikuchi, S., Yabuta, H., Nakamura, T., Tachibana, S., Ishihara, Y., Morota, T., Kitazato, K., Sakatani, N., Matsumoto, K., Wada, K., Senshu, H., Honda, C., Michikami, T., Takeuchi, H., Kouyama, T., Honda, R., Kameda, S., Fuse, T., Miyamoto, H., Komatsu, G., Sugita, S., Okada, T., Namiki, N., Arakawa, M., Ishiguro, M., Abe, M., Gaskell, R., Palmer, E., Barnouin, O.S., Michel, P., French, A.S., McMahon, J.W., Scheeres, D.J., Abell, P.A., Yamamoto, Y., Tanaka, S., Shirai, K., Matsuoka, M., Yamada, M., Yokota, Y., Suzuki, H., Yoshioka, K., Cho, Y., Tanaka, S., Nishikawa, N., Sugiyama, T., Kikuchi, H., Hemmi, R., Yamaguchi, T., Ogawa, N., Ono, G., Mimasu, Y., Yoshikawa, K., Takahashi, T., Takei, Y., Fujii, A., Hirose, C., Iwata, T., Hayakawa, M., Hosoda, S., Mori, O., Sawada, H., Shimada, T., Soldini, S., Yano, H., Tsukizaki, R., Ozaki, M., Iijima, Y., Ogawa, K., Fujimoto, M., Ho, T.M., Moussi, A., Jaumann, R., Bibring, J.P., Krause, C., Terui, F., Saiki, T., Nakazawa, S., Tsuda, Y., 2019. Hayabusa2 arrives at the carbonaceous asteroid 162173 Ryugu-A spinning top-shaped rubble pile. Science (80-. ). 364, 268–272.

Watanabe, S., Tsuda, Y., Yoshikawa, M., Tanaka, S., Saiki, T., Nakazawa, S., 2017. Hayabusa2 Mission Overview. Space Sci. Rev. 208, 3–16.

Weidenschilling, S.J., Cuzzi, J.N., 1993. Formation of Planetesimals in the Solar Nebula. In: Levy, E.H., Lunine, J.I., Guerrieri, M., Matthews, M.S. (Eds.), Protostars and Planets III. University of Arizona Press, Tucson, pp. 1689–1699.

Whiteley, R.J., Hergenrother, C.W., Tholen, D.J., 2002. Monolithic Fast-Rotating Asteroids. In: Warmbein, B. (Ed.), In: Proceedings of Asteroids, Comets, Meteors - ACM 2002. International Conference, 29 July - 2 August 2002, Berlin, Germany. ESA SP-500. Noordwijk, Netherlands: ESA Publications Division, pp. 473–480.

Wie, B., Zimmerman, B., Lyzhoft, J., Vardaxis, G., 2017. Planetary defense mission concepts for disrupting/pulverizing hazardous asteroids with short warning time. Astrodynamics 1, 3–21.

Wopenka, B., Xu, Y.C., Zinner, E., Amari, S., 2013. Murchison presolar carbon grains of different density fractions: A Raman spectroscopic perspective. Geochim. Cosmochim. Acta 106, 463–489.

Yamada, R., Senshu, H., Namiki, N., Mizuno, T., Abe, S., Yoshida, F., Noda, H., Hirata, N., Oshigami, S., Araki, H., Ishihara, Y., Matsumoto, K., 2017. Albedo Observation by Hayabusa2 LIDAR: Instrument Performance and Error Evaluation. Space Sci. Rev. 208, 49–64.

Yeomans, D.K., Barriot, J.P., Dunham, D.W., Farquhar, R.W., Giorgini, J.D., Helfrich, C.E., Konopliv, A.S., McAdams, J. V., Miller, J.K., Owen, W.M., Scheeres, D.J., Synnott, S.P., Williams, B.G., 1997. Estimating the mass of asteroid 253 Mathilde from tracking data during the NEAR flyby. Science (80-. ). 278, 2106–2109.

Zemcov, M., Immel, P., Nguyen, C., Cooray, A., Lisse, C.M., Poppe, A.R., 2017. Measurement of the cosmic optical background using the long range reconnaissance imager on New Horizons. Nat. Commun. 8, 1–9.

**Tables:**

Table 1. Initial condition of the spacecraft state in search (J2000 EQ Earth Centered Inertial frame). The initial time is 17:22:04 on December 17, 2020 (UTC).

| Components | Position [km] | Velocity [km/sec] |
|---|---|---|
| x | $3.56015012 \times 10^4$ | 0.250469472 |
| y | $-8.12279393 \times 10^4$ | $-4.43964684$ |
| z | $7.42361209 \times 10^4$ | 0.482293206 |

Table 2. Reachability Criteria (RC) derived from engineering assessment

| Items | Criteria |
|---|---|
| RC1 | Perihelion of asteroid should be less than 1.2 au. |
| RC2 | Aphelion of asteroid should be larger than 0.8 au. |
| RC3 | Maximum $\Delta V$ is less than 1.6 km/s. |
| RC4 | Spacecraft should arrive before 2032. |
| RC5 | Orbital uncertainty u code $\leq 5$. |
| RC6 | Sun-spacecraft distance less than 1.5 au when the arrival. |

Table 3. Mission Criteria (MCs) defined during mission assessment

| Items | Criteria |
|---|---|
| MC1 | Diameter: Better grade smaller asteroids. |
| MC2 | Spin period: Better grade shorter spin periods. |
| MC3 | Shape detection: Better grade asteroids with reported shapes. |
| MC4 | Material composition: Better grade reported compositions. |
| MC5 | Future observability: Better grade if there are observation chances. |
| MC6 | Orbital uncertainty: Better grade if the orbit is well determined. |
| MC7 | Existence of backups: Better grade if backup scenarios exist. |

Table 4. Final scores for determining the candidates of rendezvous targets. Column "Science only" shows the total score that only accounts for MC1 - MC4. Column "Without Backups" accounts for MC1 through MC6. Column "With Backups" shows the total score of MC1 through MC7.

| Asteroid | Science only | Without Backups | With Backups |
|---|---|---|---|
| 1998 KY26 | 5.7 | 13.7 | 13.7 |
| 2001 AV43 | 10.1 | 13.1 | 23.1 |
| 2004 VJ1 | 25. | 27. | 27. |
| 2012 MY2 | 40. | 58. | 68. |
| 2012 UW68 | 33.6 | 51.6 | 61.6 |
| 2014 JR24 | 40. | 60. | 70. |
| 2014 JV54 | 22.2 | 32.2 | 32.2 |
| 2014 YE15 | 40. | 56. | 56. |
| 2015 TJ1 | 30. | 36. | 46. |
| 2016 EP84 | 40. | 58. | 68. |
| 2016 JN6 | 32.2 | 52.2 | 52.2 |
| **2017 LD** | 40. | 58. | 68. |
| **2018 KC2** | 39.1 | 57.1 | 57.1 |
| **2019 JU5** | 35.2 | 39.2 | 49.2 |

Table 5. Time schedule of EVEEA scenario

| Time | Event | Scientific observations |
|---|---|---|
| **2021-2024** | Long-term cruise | Zodiacal light observations<br>Exoplanet observations |
| **2024** | Venus flyby | Venus atmosphere observations |
| **2025** | Earth swing-by 1 | Calibrations of scientific instruments |
| **2026** | Earth swing-by 2 | Calibrations of scientific instruments |
| **2029** | Rendezvous with 2001 AV43 | Proximity observations of a small asteroid |

Table 6. Time schedule of E(A)EEA scenario

| Time | Event | Scientific observations |
|---|---|---|
| **2021-2026** | Long-term cruse | Zodiacal light observations<br>Exoplanet observations |
| **2026** | 2001 CC21 flyby | Fly-by observations by using remote sensing instruments. |
| **2027** | Earth swing-by 1 | Calibrations of scientific instruments |
| **2028** | Earth swing-by 2 | Calibrations of scientific instruments |
| **2031** | Rendezvous with 1998 KY26 | Proximity observations of a small asteroid |

Table 7. Physical properties of 2001 AV43.

| Physical Properties | Conditions |
|---|---|
| **Shape** | Elongated (Aspect ratio, ~0.5) |
| **Effective diameter [m]** | ~40 |
| **Spin period [min]** | 10.2 |
| **Non-principal components of rotation** | Not observed |
| **Taxonomy** | Not well known (possibly S) |

Table 8. Physical properties of 1998 KY26

| Physical Properties | Conditions |
|---|---|
| **Shape** | Relatively spherical (see Figure 7) |
| **Effective diameter [m]** | 20 – 40 |
| **Spin period [min]** | $10.7015 \pm 0.0004$ |
| **Non-principal components of rotation** | Not observed |
| **Taxonomy** | B, C, F, G, D, or P |

Table 9. Traceability matrix for proximity operations. MO stands for the mission objectives (Section 4). SO gives the science objectives that address issues defined by a given MO. S/C stands for Spacecraft.

| MO ID | SO ID | Items | Investigations | Measurement requirements |
|---|---|---|---|---|
| 2, 3 | 1 | Identify whether the asteroid is rubble pile or monolithic. | Observe the existence of smaller particles and surface morphologies. | Global mapping using ONC-T with a resolution of 1 cm. |
| 2, 3 | 2 | Constrain the surface evolution processes. | Measure the spectral signature, depending on the material distributions. | Global and local mapping using NIRS3 and ONC-T with multiple filters |
| 2 | 3 | Determine the environment surrounding the target body. | Search for particles flying around the target body. | Use ONC-T/TIR/LIDAR to conduct global search. |
| 3 | 4 | Characterize the surface material conditions including strength and porosity. | Measure the temperature variation and thermal inertia distribution. | Global mapping using TIR. |

Table 10. Physical properties of 2001 CC21.

| Physical Properties | Conditions |
|---|---|
| Shape | Unknown |
| Effective diameter [m] | ~700 m (Assuming a geometric albedo of 0.15) |
| Spin period [h] | ~5 |
| Non-principal components of rotation | No observed |
| Taxonomy | Possibly L |

**Figure:**

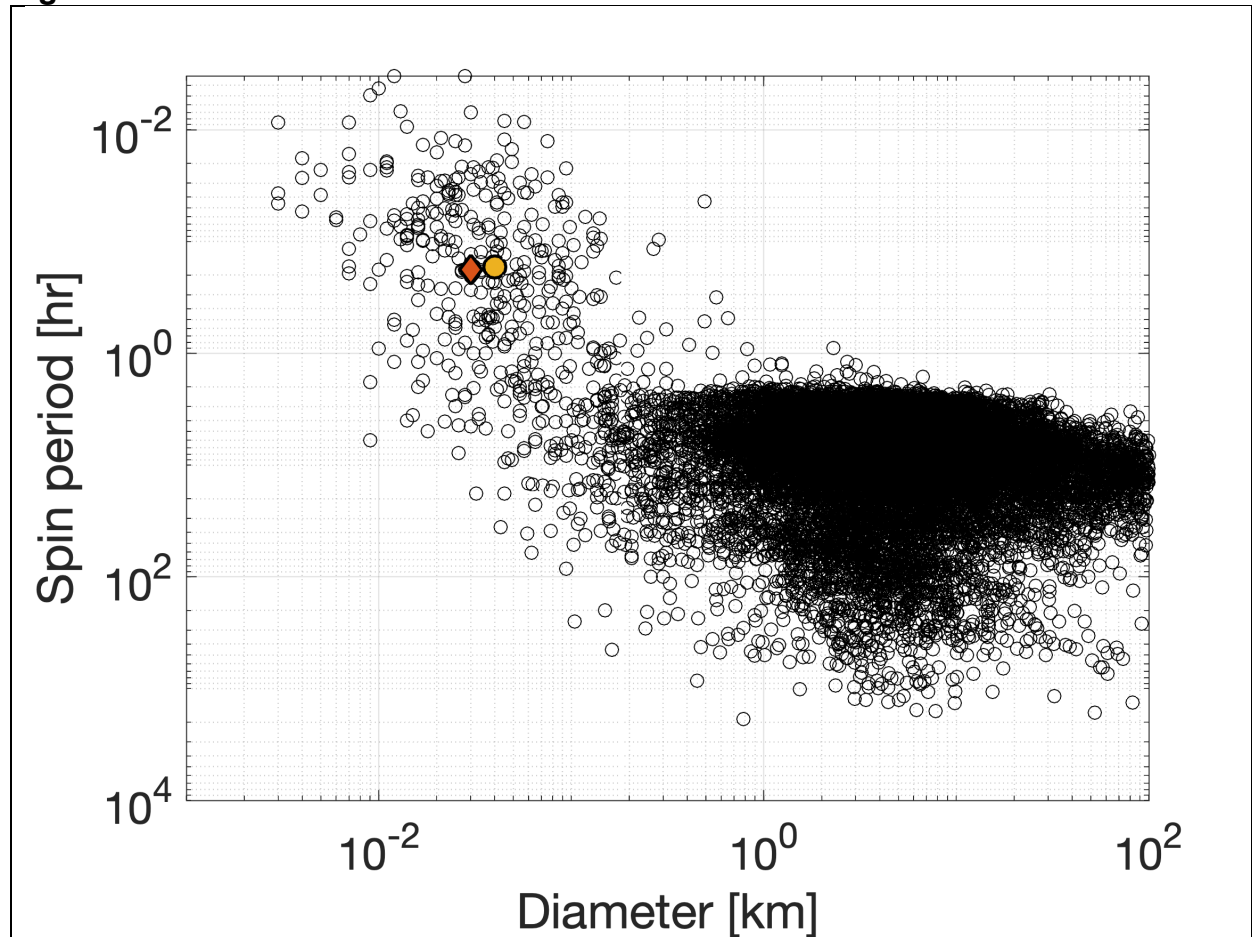

Figure 1. Dependence of asteroids' spin rates on their diameters (Warner et al., 2009). Plotted are asteroids with U-codes equal to or higher than 2. Black circles with empty space show orbital groups/families (near-Earth objects, Main Belt, Eos, Erigone, Eunomia, Flora, Koronis, Nysa, and Phocaea). The diamond filled with red shows 1998 KY26, and the circle filled with orange is 2001 AV43.

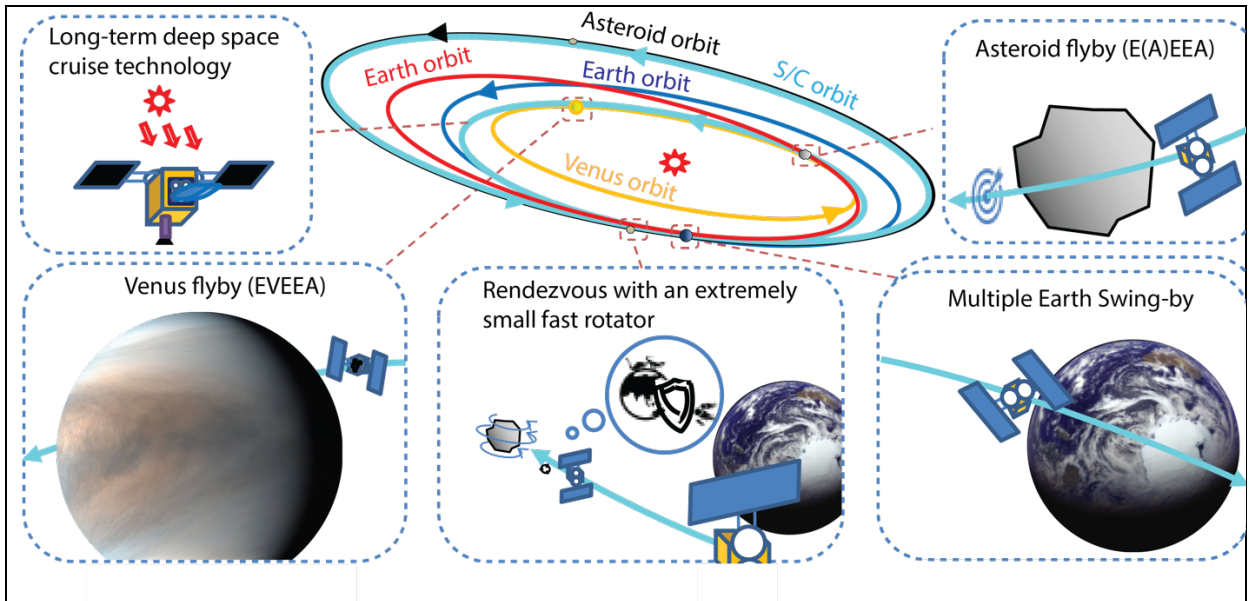

Figure 2. Schematic of the planned extended mission. Credits of images of Earth and Venus: JAXA.

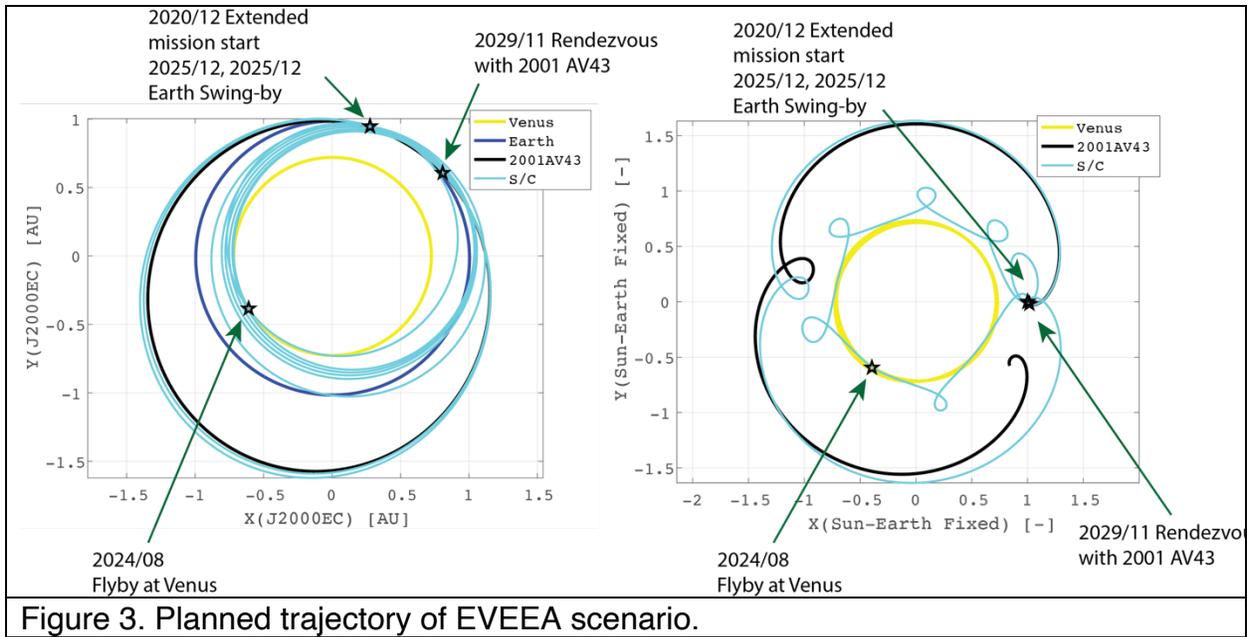

Figure 3. Planned trajectory of EVEEA scenario.

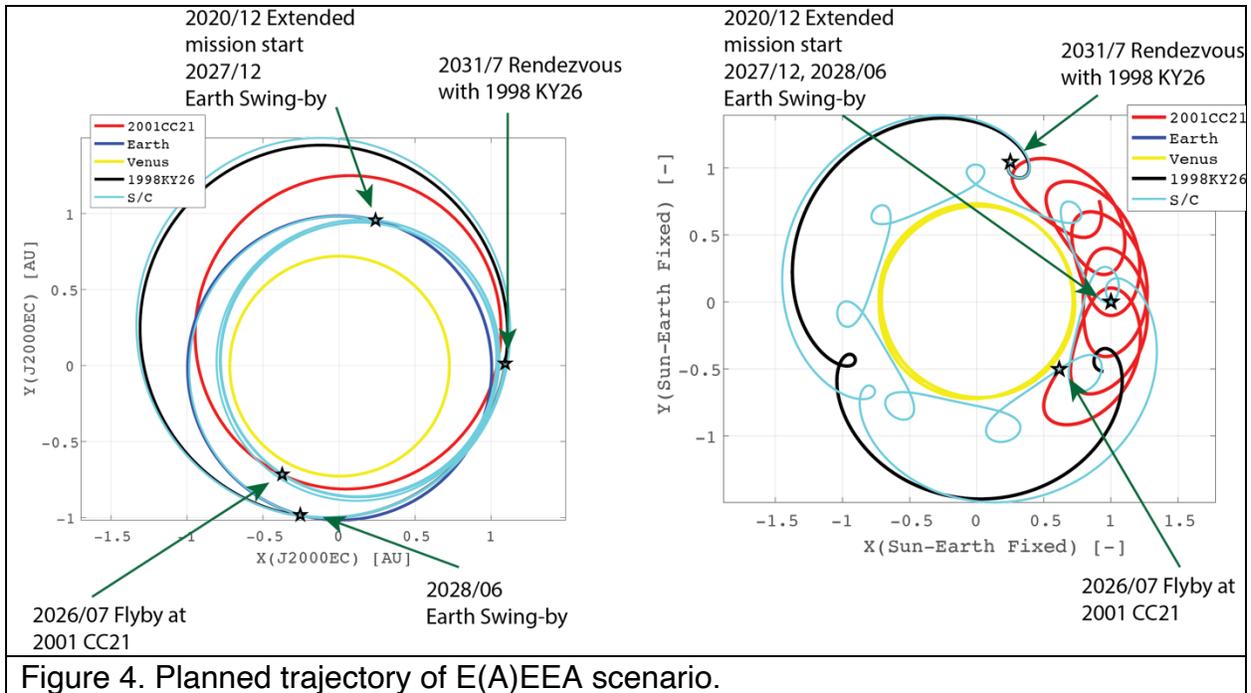

Figure 4. Planned trajectory of E(A)EEA scenario.

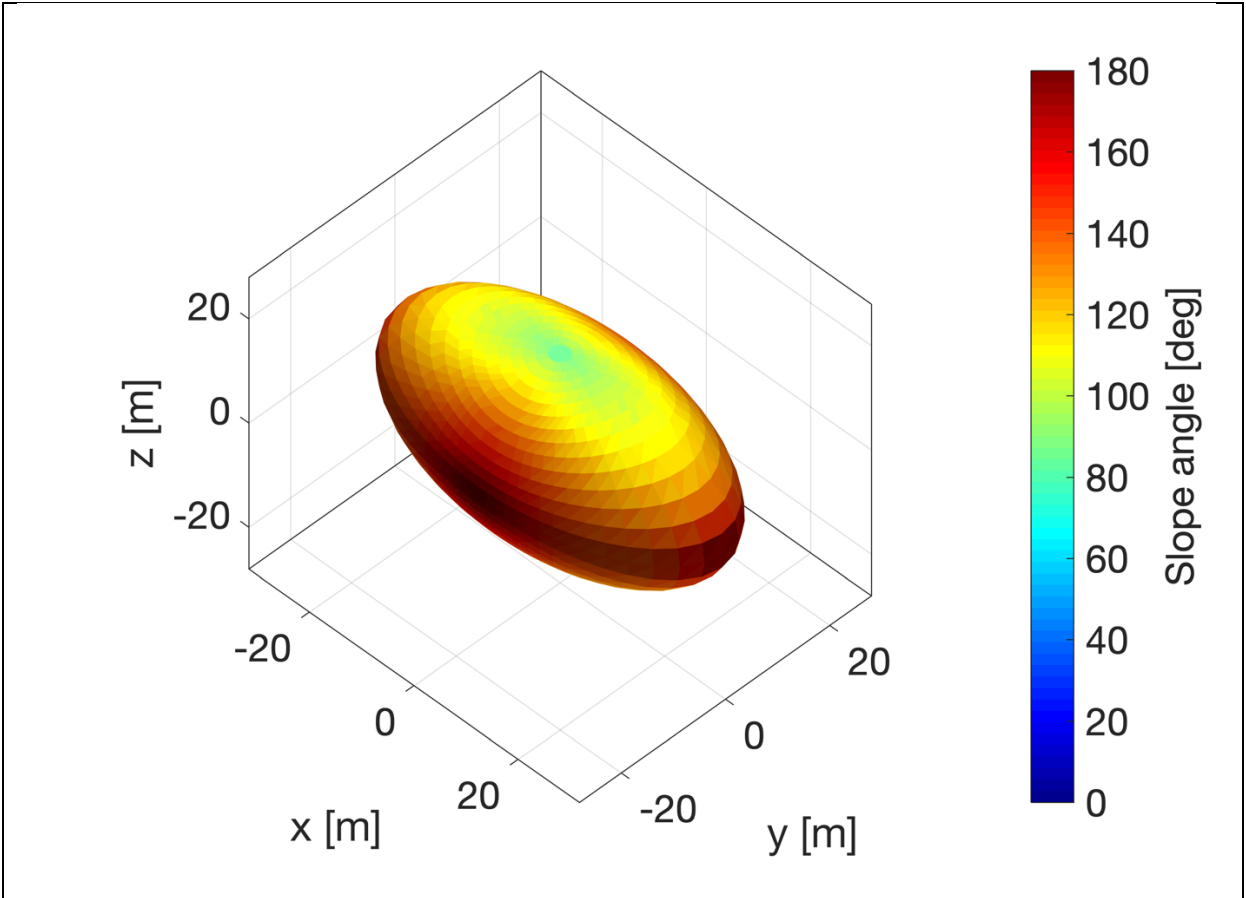

Figure 5. Gravitational slope of 2001 AV43. The shape was assumed to be a triaxial ellipsoid with an aspect ratio of 0.5.

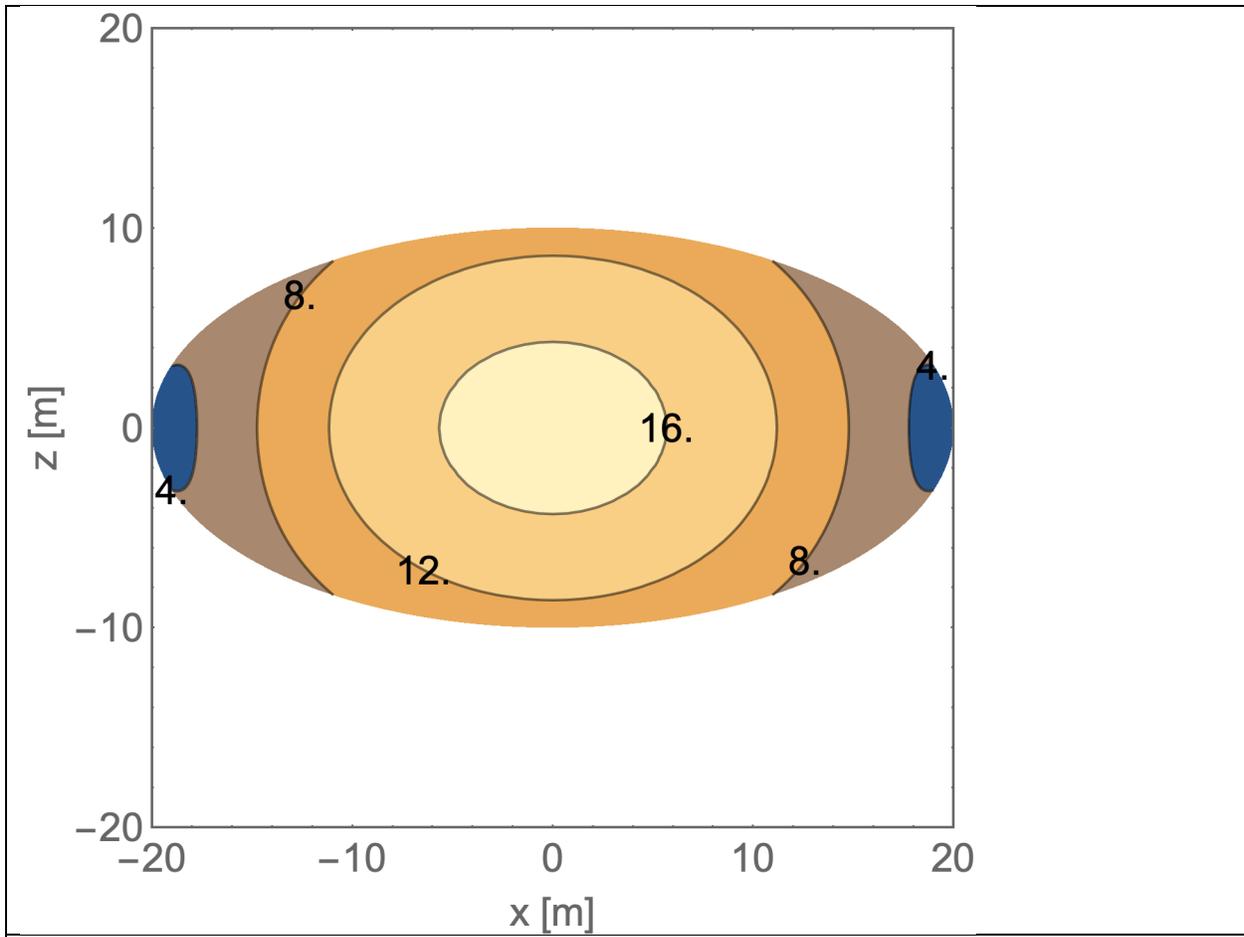

Figure 6. Minimum cohesive strength in 2001 AV43. The $x$ axis is along the longest axis, and the $z$ axis is along the spin axis.

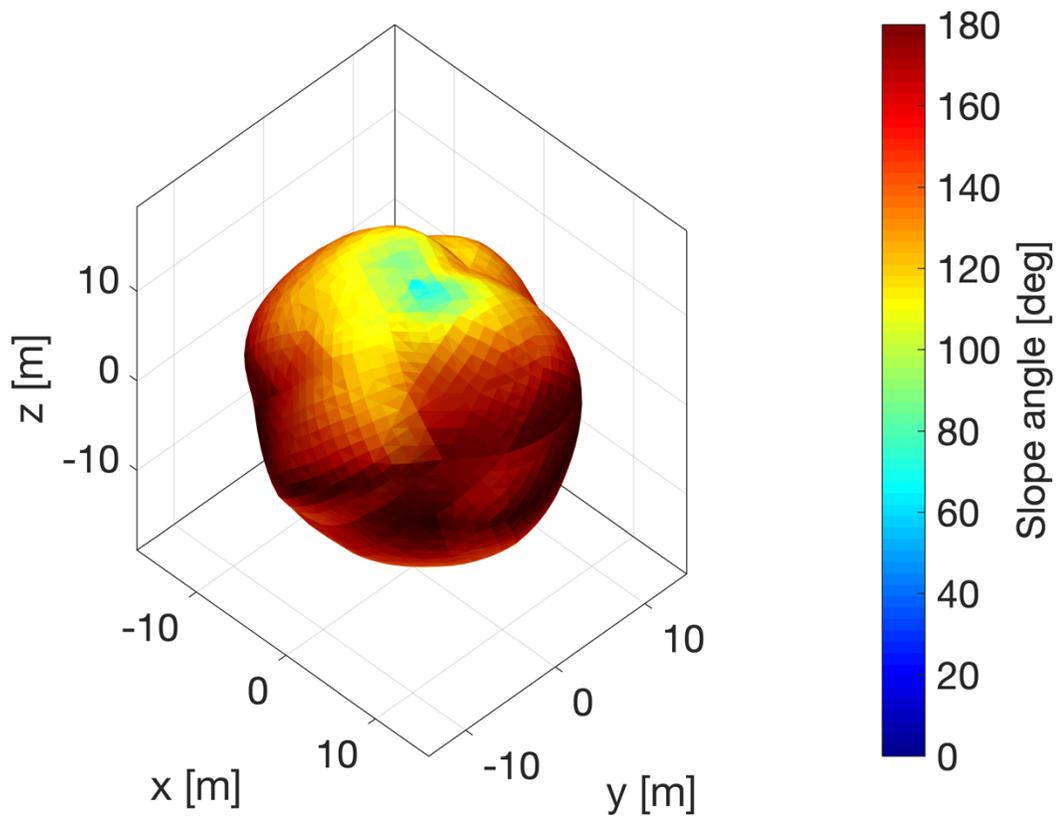

Figure 7. 1998 KY26's shape model from least-squares fitting to the radar echo spectra (Ostro et al., 1999).

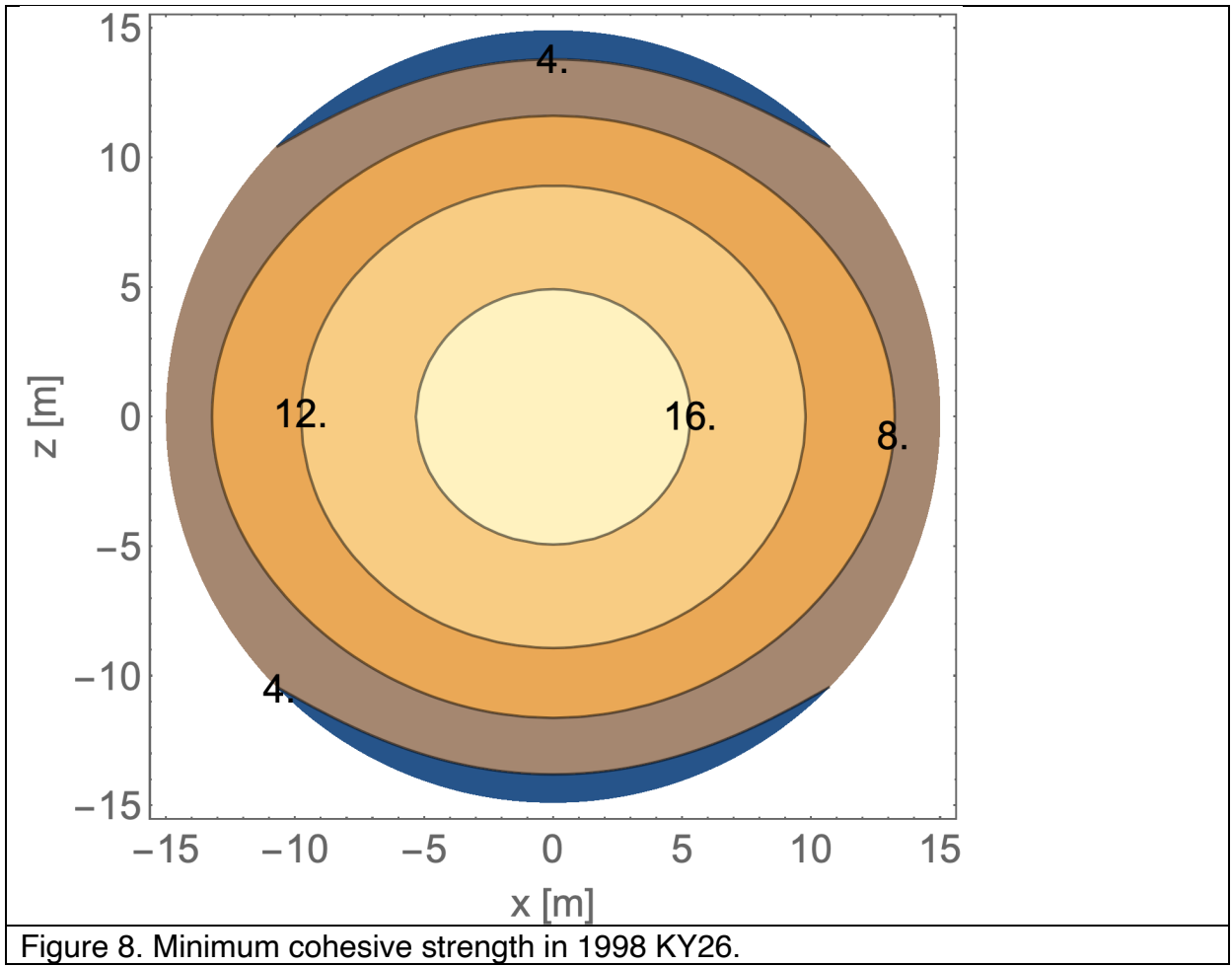

Figure 8. Minimum cohesive strength in 1998 KY26.

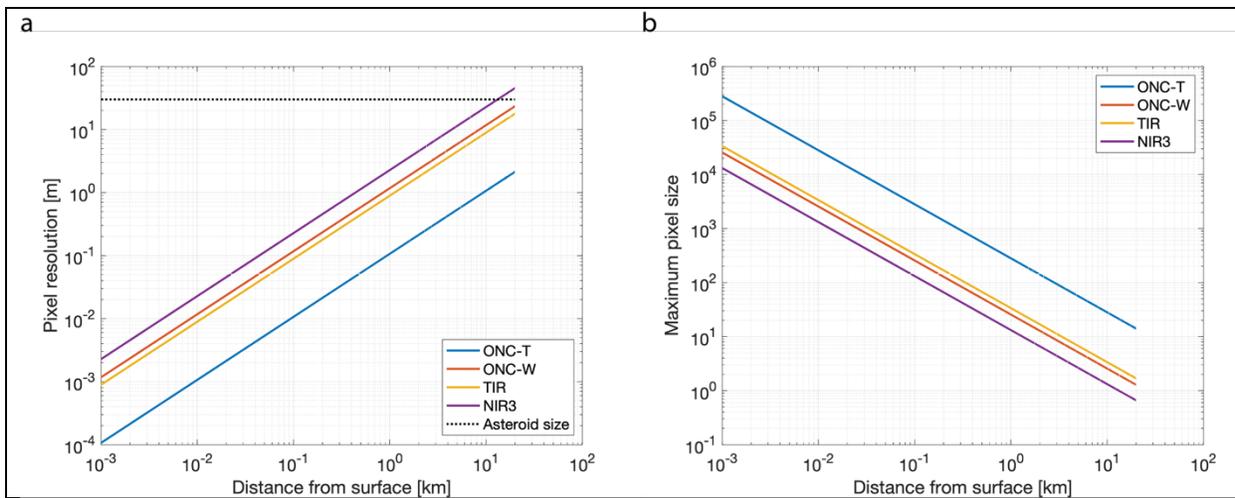

Figure 9. Relationships between asteroid size and onboard instrument spatial resolutions. a. Pixel resolution of remote sensing instruments. b. Maximum pixel resolution.